\documentclass[12pt]{article}

\usepackage[top=2.5cm,bottom=2.5cm,left=1.0cm,right=1.0cm,headsep=10pt,a4paper]{geometry} 

\usepackage{authblk}
\usepackage{graphicx}
\usepackage{dcolumn}
\usepackage{bm}
\usepackage[mathlines]{lineno}
\usepackage{latexsym}
\usepackage{amssymb}
\usepackage{amsmath}
\usepackage{epstopdf}
\usepackage{eurosym}

\usepackage[nottoc,numbib]{tocbibind}

\usepackage{xcolor} 
\usepackage[colorinlistoftodos]{todonotes}

\newcommand{\ket}[1]{|#1\rangle}

\newcommand{\bc}{\begin{cases}\begin{aligned}}
\newcommand{\ec}{\end{aligned}\end{cases}}
\newcommand{\uno}{\leavevmode\hbox{\small1\normalsize\kern-.33em1}}
\newcommand{\beq}{\begin{equation}}
\newcommand{\eeq}{\end{equation}}

\title{SAGE: A Proposal for a Space Atomic Gravity Explorer}

\author[1]{G. M. Tino}
\author[2]{A. Bassi}
\author[3]{G. Bianco} 
\author[4]{K. Bongs} 
\author[5]{P. Bouyer} 
\author[6]{L. Cacciapuoti}
\author[7]{S. Capozziello} 
\author[8]{X. Chen} 
\author[9]{M. L. Chiofalo}
\author[10]{A. Derevianko} 
\author[11]{W. Ertmer} 
\author[11]{N. Gaaloul}
\author[12]{P. Gill} 
\author[13]{P. W. Graham} 
\author[13]{J. M. Hogan}
\author[14]{L. Iess }
\author[13]{M. A. Kasevich} 
\author[15]{H. Katori} 
\author[11]{C. Klempt} 
\author[16]{X. Lu}
\author[17]{L.-S. Ma }
\author[18]{H. M\"uller} 
\author[19]{N. R. Newbury} 
\author[19]{C. Oates} 
\author[20]{A. Peters}
\author[1]{N. Poli}
\author[11]{E. Rasel}
\author[1]{G. Rosi} 
\author[21]{A. Roura} 
\author[22]{C. Salomon} 
\author[23]{S. Schiller} 
\author[21]{W. Schleich} 
\author[11]{D. Schlippert} 
\author[24]{F. Schreck} 
\author[11]{C. Schubert} 
\author[25]{F. Sorrentino} 
\author[26]{U. Sterr}
\author[27]{J. W. Thomsen} 
\author[28]{G. Vallone}  
\author[29]{F. Vetrano} 
\author[28]{P. Villoresi} 
\author[30]{W. von Klitzing} 
\author[31]{D. Wilkowski}
\author[32]{P. Wolf} 
\author[33]{J. Ye}
\author[34]{N. Yu} 
\author[35]{M. S. Zhan}

\affil[1]{Dipartimento di Fisica e Astronomia and LENS, Universit\`{a} di Firenze, INFN, via Sansone 1, I-50019 Sesto Fiorentino (Firenze), IT}
\affil[2]{Department of Physics, University of Trieste, IT}
\affil[3]{Space Geodesy Centre, ASI, Matera, IT}
\affil[4]{School of Physics and Astronomy, University of Birmingham, UK}
\affil[5]{Laboratoire Photonique, Numerique et Nanosciences, Bordeaux, FR}
\affil[6]{European Space Agency, Noordwijk ZH, NL}
\affil[7]{Dipartimento di Fisica, INFN, Universit\`{a} di Napoli ``Federico II", IT}
\affil[8]{Peking University, Beijing, China }
\affil[9]{Dipartimento di Fisica, INFN, Universit\`{a} di Pisa, IT} 
\affil[10]{Physics Department, University of Nevada, Reno, USA }
\affil[11]{Institute of Quantum Optics and QUEST-Leibniz Research School, Leibniz Universit\"at Hannover, Hannover, DE}
\affil[12]{National Physical Laboratory, Teddington, Middlesex,  UK}
\affil[13]{Stanford University, Stanford, CA, USA }
\affil[14]{Dipartimento di Ingegneria Meccanica e Aerospaziale,  Universit\`{a} di Roma, IT}
\affil[15] {University of Tokyo, RIKEN, Tokyo, JP}
\affil[16]{Zhejiang University, Hangzhou, CN}
\affil[17]{East China Normal University, Shanghai, CN}
\affil[18]{University of California, Berkeley, CA, USA}
\affil[19]{NIST, Boulder, Colorado, USA}
\affil[20]{Humboldt-Universit\"at zu Berlin, DE}
\affil[21]{Institut f\"ur Quantenphysik, Universit\"at Ulm, DE}
\affil[22]{Laboratoire Kastler Brossel, Ecole Normale Sup\'erieure, Paris, FR}
\affil[23]{Institut f\"ur Experimentalphysik, Heinrich-Heine-Universit\"at D\"usseldorf, DE} 
\affil[24]{Van der Waals-Zeeman Institute, University of Amsterdam, NL}
\affil[25]{Istituto Nazionale di Fisica Nucleare, Sezione di Genova, IT} 
\affil[26]{Physikalisch-Technische Bundesanstalt (PTB), Braunschweig, DE}
\affil[27]{Niels Bohr Institute, Copenhagen, DN}
\affil[28]{Department of Information Engineering, University of Padova, IT} 
\affil[29]{DiSPeA, University of Urbino Carlo Bo, IT} 
\affil[30]{Institute of Electronic Structure and Laser, Foundation for Research and Technology-Hellas, Heraklion 70013, GR} 
\affil[31]{School of Physical and Mathematical Sciences, Nanyang Technological University, 637371 Singapore, SGP}
\affil[32]{LNE-SYRTE, Observatoire de Paris, CNRS, PSL, Sorbonne Universit\'e, FR} 
\affil[33]{JILA, NIST, and University of Colorado, Boulder, Colorado, USA}
\affil[34]{Jet Propulsion Laboratory, California Institute of Technology 4800 Oak Grove Ave., Pasadena, CA  91109, USA}
\affil[35]{Wuhan Institute of Physics and Mathematics, CAS, Wuhan, CN}

\begin{document}

\maketitle

%

%


\begin{abstract}
The proposed mission "Space Atomic Gravity Explorer" (SAGE) has the scientific objective to investigate 
 gravitational waves, dark matter, and other fundamental aspects of gravity as well as the connection between gravitational physics and quantum physics using new quantum sensors, namely, optical atomic clocks and atom interferometers based on ultracold strontium atoms. 
\end{abstract}

\section{Introduction}{\label{Introduction}}

The SAGE mission was proposed to the European Space Agency in 2016 in response to a Call for ``New Ideas" \cite{ESA2016}.  Combining quantum sensing and quantum communication, SAGE is based on recent impressive advances in quantum technologies for atom interferometers \cite{Cronin2009,Tino2014}, optical clocks \cite{Poli2013,Ludlow2015}, and microwave and optical links, which now enable new high-precision tests in fundamental physics \cite{Safronova2018}.  This paper describes  a  multi-purpose gravity explorer mission based on the most advanced achievements in the field.

SAGE's main goals are:

\begin{itemize}
\item  Observing gravitational waves in new frequency ranges with atomic sensors.
\item  Searching for dark matter.
%
%
\item  Testing general relativity and the Einstein equivalence principle.
\item  Investigating quantum correlations and testing Bell inequalities for different gravitational potentials and relative velocities.
\item  Defining an ultraprecise frame of reference for Earth and Space, comparing terrestrial clocks, using clocks and links between satellites for optical VLBI in Space.

\end{itemize}

We consider a multi-satellite configuration with payload/instruments including strontium optical atomic clocks, strontium atom interferometers, satellite-to-satellite and satellite-to-Earth laser links.

Although the technology for the SAGE mission is not completely mature yet, it takes advantage of developments for the ACES (Atomic Clock Ensemble in Space)  \cite{Cacciapuoti2009,Laurent2015} and CACES (Cold Atom Clock Experiment in Space) \cite{Liu2018} missions and the results of ESA studies for SOC (Space Optical Clock) \cite{Bongs2015,ISOC-ESR,Origlia2018}, SAI (Space Atom Interferometer) \cite{Tino2007a,Sorrentino2010}, SAGAS (Search for Anomalous Gravitation using Atomic Sensors) \cite{Wolf2009}, STE-QUEST (Space-Time Explorer and QUantum Equivalence principle Space Test) \cite{Aguilera2014,Altschul2015}, and other ongoing  or planned projects on ground \cite{Canuel2018,MAGIS100,Zhan2019} and in space \cite{Becker2018,GOAT,Gao2018,Wang2019}.

This paper is organized as follows: for the main goals of the SAGE mission, Section 2 briefly introduces the science case, in Section 3  the scientific requirements are discussed in  detail, Section 4 describes the measurement concept. Finally, Section 5 provides preliminary information on the present technology readiness level and possible roadmap towards the mission.


\section{\label{Science case} Science case and mission objectives }

Recent advances in quantum technologies based on atomic physics  and optics have provided new tools for the experimental investigation of gravity.  These tools include atom interferometry, where quantum mechanical interference of atomic de Broglie waves enables extremely precise measurements, the new ultra-precise optical atomic clocks, and ultra-stable lasers.  The proposed combination of these methods in SAGE based on ultracold strontium (Sr) atoms enables a new generation of ultra-precise sensors suited to gravitational wave (GW) detection in unexplored frequency ranges, search for dark matter (DM), and to other fundamental tests of gravitational physics in space, with extreme precision.

\subsection{Observing gravitational waves in new frequency ranges with atomic sensors \label{GWs:science} }


The detection of  GWs by LIGO/Virgo optical interferometers \cite{abbott2016observation, abbott2017b} at frequencies from tens to hundreds of Hz opened a new window for observing the Universe. In the improved configuration terrestrial detectors are expected to observe tens of GW events per year in this frequency range. At the same time, there is a large interest in extending the range of GW detectors towards low frequencies, where a much larger number of observable sources is predicted. The successful operation of LISA Pathfinder \cite{Armano2016} paves the way to the space optical detector eLISA \cite{L3_final_report}, which will observe GWs at low frequencies, with a peak sensitivity around $1 - 10$~mHz. 

The basic idea of the SAGE proposal is to use atomic sensors, namely atomic clocks and atom interferometers based on ultracold strontium atoms, for the observation of gravitational waves in the low frequency range from $10^{-3}$~Hz to $10$~Hz. This will complement eLISA with an alternative technology, and more importantly, it will fill the gap between the eLISA and terrestrial detectors.

The possibility of using atom interferometers to detect gravitational waves has been investigated for some time \cite{Tino2007b,Tino2011,Dimopoulos2008}. Schemes based on optical atomic clocks were  proposed recently \cite{Kolkowitz2016}. Both ideas can be implemented with the same atomic species, namely, strontium. It is already used in optical clocks reaching $10^{{-18}}$ uncertainty and instability, and also has the desired properties for advanced atom interferometry \cite{Ferrari2006b,Mazzoni2015,Hu2017,Loriani2019}. It is worth noting 
\cite{Norcia2017} that at a fundamental level atomic clocks and atom interferometers have the same mechanism for sensitivity. They both sense changes in phase that result from changes in optical path length between the two satellites. The main difference in the two schemes is how the atomic phase reference is interrogated: either with atoms in free fall (atom interferometers) or confined in optical lattices (atomic clocks)
\cite{Norcia2017}.
The main advantages of GW detectors in space are the absence of major seismic and gravity gradient noises, which prevent the observation of low-frequency GW signals by ground-based instruments, and the much larger baseline, which allows to enhance the strain sensitivity. Atomic sensors are particularly well suited for the precise measurement of long-term phase and frequency changes of optical fields; indeed atoms are at the same time excellent test masses, which can be precisely manipulated with optical fields, and excellent clocks, providing ultra-stable  frequency references. In addition atomic clocks and interferometers are extremely flexible, allowing to dynamically adapt the spectral sensitivity by tuning simple experimental parameters.  As we show in the following sections, this suggests interesting complementarities between the atomic and the optical GW detectors.

The GW spectral range that SAGE is aiming to investigate is extremely interesting since it is expected that several GW sources can be observed for a long time, resulting in an increase of their detectability. While recent findings suggest that most stars are single \cite{Lada2006}, at least one third of them are in binary or higher multiple systems. Binary systems are believed to be precursor to various interesting phenomena as for example the formation of neutron stars, type Ia supernovae and the evolution of supermassive black holes in galactic nuclei. The GW signals emitted by binary systems before the merger are characterized by a long Newtonian inspiralling phase, well modelled by simple analytical expressions for objects like neutron stars, black holes and white dwarfs.

The GW events observed by LIGO/Virgo detectors \cite{abbott2016observation, abbott2017b} suggest that there may be a large population of stellar-mass black-hole/black-hole binary systems of significant total mass ($>50\,M_{\odot}$).
A space detector could be able to resolve these sources at modest red-shift, and to observe them until the coalescence becomes detectable by ground-based antennas, with all the benefits of a joint longer observation. A re-evaluation of the science reach of a space detector will occur as soon as the terrestrial LIGO-Virgo network will detect a number of neutron-star/neutron-star and neutron-star/black-hole systems, following the outstanding event in 2017 \cite{Abbott2017observation}.

\begin{figure}
    \centering
\includegraphics[width=0.6 \columnwidth]{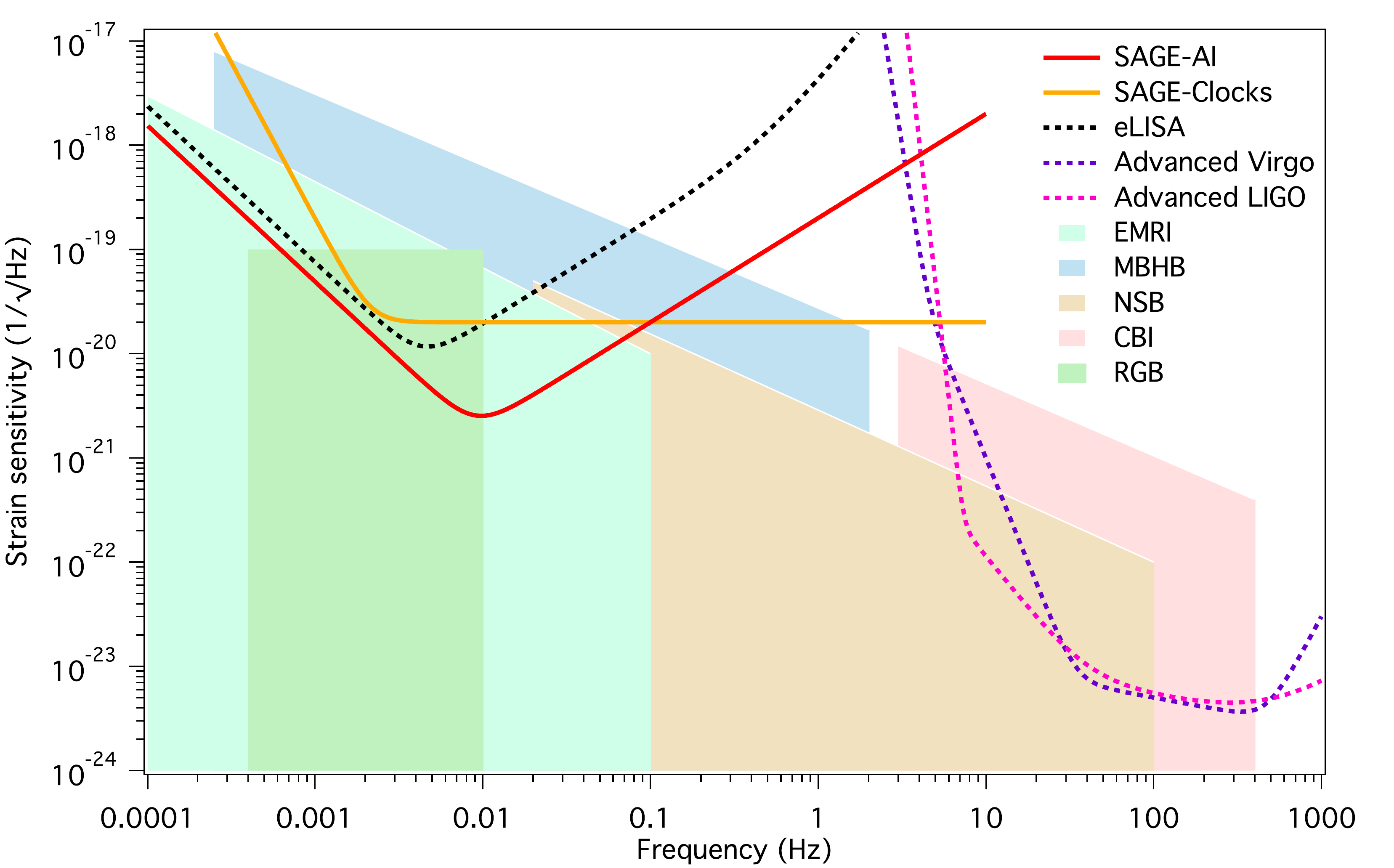}
    \caption{Spectral strain sensitivity of the SAGE GW detector for the two proposed schemes described in Sect. \ref{GW_concept}. 
    SAGE-AI refers to the atom interferometry scheme with 12 $\hbar k$ atomic beam splitters  (Sect. \ref{AI_concept}).
The SAGE-Clocks curve represents the optimized performance of the tuned detector for the scheme using optical clocks (Sect. \ref{clock_concept}).
    The design sensitivity of eLISA \cite{L3_final_report}, Advanced Virgo \cite{AdV} and Advanced  LIGO \cite{AdL}  is reported for comparison. Shaded areas represent  the expected spectral strain amplitude of GWs from  sources within the accessible frequency ranges of future space detectors and current ground detectors.
 EMRI: extreme mass ration inspirals; MBHB: massive black-hole binaries; NSB: neutron star binaries; CBI: compact binary inspirals; RGB: resolved galactic inspirals.}
    \label{fig.sources_LISA}
\end{figure}

A summary of potentially accessible GW sources is shown in Fig.~\ref{fig.sources_LISA}.
Massive black-holes are supposed to be in the center of galaxies: massive black-hole binaries (MBHBs) trace the ongoing mergers of these galaxies when they are merging, supplying us with a lot of information about the role of these black holes on galaxy formation and evolution. Furthermore these systems are expected to be some of the strongest sources for detectable low-frequency gravitational waves, providing what can be considered a kind of gold-plated event of gravitational
astronomy. In many cases, all the relevant parameters of the system (masses, spin, distance and sky location) can be well determined from the properties of the observed gravitational wave.
Extreme-mass-ratio inspirals (EMRIs) are systems in which a body orbits a much more massive body, with emission of gravitational waves that leads to a gradual orbit decay. Very interesting examples of these systems are supposed at the center of galaxies, where stellar black holes and neutron stars may orbit a supermassive black hole. The slow evolution of these systems implies many thousands of cycles before eventually plunging:  the  gravitational wave signal enables a precise mapping of the spacetime geometry of the supermassive black hole. It is expected that a space detector with peak sensitivity better than $10^{-20}$ would be able to detect thousands of EMRIs and allow the determination of the system parameters.                                                                                                       
Several millions of white dwarf binaries in our Galaxy are predicted to emit gravitational waves in the $10^{-3} - 10$\,Hz band, generating  a stochastic GW background. Some of them are known from electromagnetic observations, and might serve as Ôverification binariesÕ. A space detector with $10^{-20}$ peak strain sensitivity would be capable to resolve several thousands of such sources. Each signal, almost monochromatic (apart from  a small drift in frequency due to gravitational radiation and/or to mass transfer) lasts a long time, enabling a good characterisation of the system by a complete set of astrophysical  parameters (seven or eight, depending on the capability in measuring the frequency drift \cite{shah2014}) . 

SAGE will employ  atom interferometers and optical atomic clocks as novel detectors for these GW sources, complementary to optical interferometers on Earth (LIGO, Virgo) and planned for Space (LISA, eLISA) \cite{L3_final_report}.
Atomic detectors can indeed benefit from quantum metrology methods that can be used to achieve optimal, frequency-independent  sensitivity to GWs in the whole measurement range of interest \cite{Kolkowitz2016}. In particular, we consider two possible measurement schemes, based on the use of  ultracold Sr atoms to be employed  either as test masses  in light-pulse atom interferometry or as frequency references for optical atomic clocks. Potential advantages of the use of atomic sensors, which are common to the two approaches, include the tunability of the sensitivity curve, quantum back-action noise immunity, and insensitivity to laser frequency noise, which in turn  allows for the possibility of a detector design based on a single linear baseline, requiring only two satellites instead of three. 

Further advantages of the proposed idea include the possible phase multiplication using multiple-pulse sequences, and the resilience of the proof mass: the properties of the atomic proof-masses are indeed regular  and well-known. Moreover, these test masses are virtually immune from several spurious effects such as charging events. 
The quantum nature of atomic sensors, which are  based on resonant atom-light interaction, relaxes the requirements on transmitted laser power, thus allowing longer baselines.  The frequency measurement scheme  would allow a continuous tuning of the sensitivity curve in order to observe a GW signal from its first detection through its evolution up to the frequency range where terrestrial detectors can start to follow the signal. This scheme would allow to increase the baseline, and thus to improve the low-frequency sensitivity, without sensitivity degradation at higher frequencies.


%

Differently from other space and Earth-based GW  observatories, the SAGE mission will have a multi-purpose application to different experiments in fundamental physics, as discussed in the following.

\subsection{Searching for Dark Matter }\index{DM:Science}
A variety of cosmological-scale observations (e.g., gravitational lensing, galactic rotation curves,  peaks in the cosmic microwave background spectra)  indicate that ordinary visible or baryonic matter makes up only $5 \%$ of the total energy density of the universe, with the remaining balance attributed to dark matter (DM) and dark energy. What is the microscopic composition of DM? Are there non-gravitational interactions of DM with standard model particles and fields? Given these interactions, what are the strategies enabling direct detection of DM?  These are open questions and  major challenges to the present state of knowledge in both theoretical and experimental physics. 

Most of the direct DM  searches focus  on heavy DM particles, with mass much larger than an eV, looking for energy deposition  from DM particle scattering in detectors~\cite{Feng2010,Akerib:2013tjd}. However, even for particle DM candidates, the masses can be as low as  $10^{-22} \, \mathrm{eV}$, 
a range that 
can be  extended further if such particles do not make up all of the 
DM~\cite{Marsh2016}.  Recent theoretical work~\cite{Hu2000,Marsh2016,Marsh2014,Hui2017}  on ``fuzzy dark matter'' focuses on  the $10^{-22}-10^{-21}\, \mathrm{eV}$ range and the ``axiverse'' candidates~\cite{Arvanitaki2010} extend the lower mass bound all the way down to $10^{-33} \, \mathrm{eV}$. Traditional particle detection techniques are limited by their energy thresholds, making it challenging to search for lighter DM candidates: new types of technology are required to search for such ultralight DM candidates. We refer the reader to ~\cite{Safronova2018} for a recent review on ultralight DM searches.

\begin{figure}
    \centering
    \fbox{\includegraphics[width=0.8 \columnwidth]{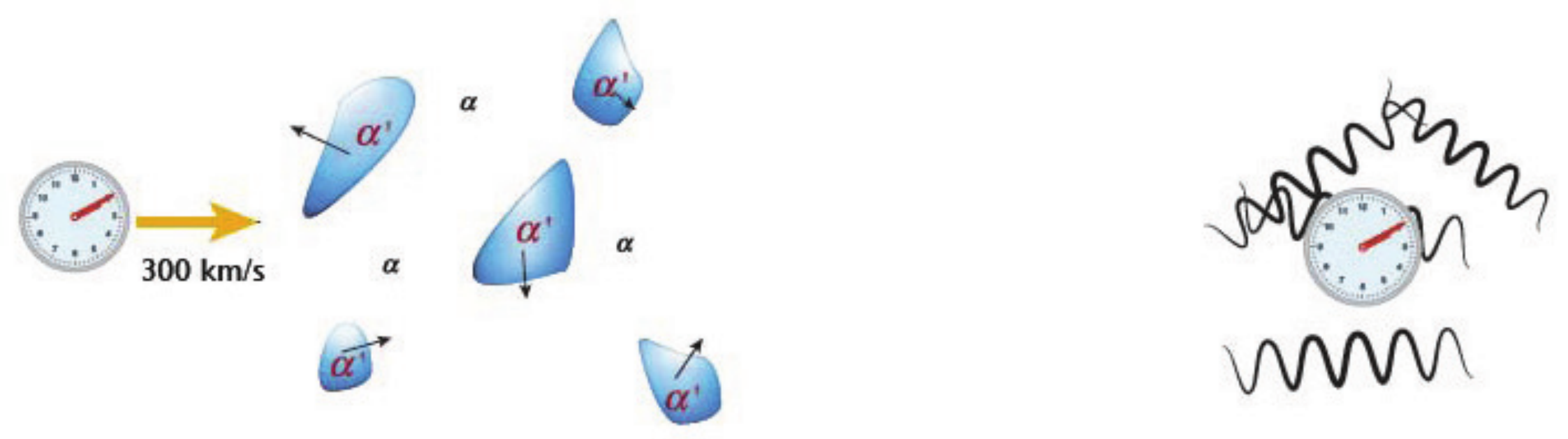}}
    \caption{  Searching for DM with atomic clocks. (Left) An atomic clock sweeps through DM constituents at galactic velocity. DM is considered to be composed of clumps. The clumps can  slow down or speed up the clock 
    tick rate
    if 
    the values of the fundamental constants (e.g. the fine-structure constant $\alpha$) differ outside and inside  the clumps  \cite{Derevianko2014}. (Right)  An oscillation of the value of fundamental constants at the field's Compton frequency can be produced by ultra-light fields. One can search for DM by Fourier analysis of clock frequency measurements and looking for peaks in the  spectrum
     \cite{Arvanitaki2015}. Figure taken with modifications from \cite{Derevianko2016} (https://creativecommons.org/licenses/by/3.0/).}
    \label{fig.DM}
\end{figure}

Several papers have recently reported  limits on dark matter models~\cite{Tilburg2015, Hees2016, Hees2018, Wcislo2018, Roberts2017-GPS-DM}. We propose to use the SAGE configuration of high-precision Sr optical atomic clocks and atom interferometers in space with optical links for new searches of ultralight DM fields. Different  models lead to DM in the form of extended clumps or oscillating fields. DM can then be detected as transient effects when the atomic sensors travel through the DM halo or as oscillating effects at Compton frequencies for non-interacting fields (Fig.~\ref{fig.DM}). For the SAGE mission, the DM constituents must have interactions with standard model fields and particles such that the fundamental constants are affected. Thereby, DM search can be viewed as a search for spatio-temporal variation of fundamental constants that is consistent with the standard halo model of dark matter distribution in our galaxy. Such fundamental constants variations are registered as transients for clumpy DM~\cite{Derevianko2014}  and as oscillations (albeit stochastic~\cite{Derevianko2016a}) for non-interacting fields~\cite{Arvanitaki2015}.

SAGE has two spatially separated atomic sensors and our proposal takes  advantage of such 
a
two-node network. DM search strategies with a geographically distributed network of precision sensors can be found in~\cite{Derevianko2014} for clumply DM and in Ref.~\cite{Derevianko2016a} for non-interacting fields. While both strategies can be pursued by SAGE, here we focus on the transients. The relevant experimental constraints on transients come from atomic clocks on board GPS~\cite{Roberts2017-GPS-DM} and 
from 
laboratory clocks~\cite{Wcislo-clock-network-2018,Wcislo2016}. 
SAGE can contribute to the search of DM by extending the Earth-based network of quantum sensors. The distance between SAGE and Earth can be important for distinguishing spurious from real detection signals.

\subsection{Testing General Relativity and the Einstein Equivalence Principle \label{EEP:Science}}

General Relativity (GR) and all metric theories of gravitation are founded on the so called  Einstein Equivalence Principle (EEP). This principle plays a major role in physics today and is therefore, not surprisingly, the subject of ever more precise experimental scrutiny. 
Additionally, most unification theories (theories that try to unify GR and the 
Standard Model 
of particle physics) have in common a violation of the EEP at some level which is unknown a priori. Since the EEP is not a fundamental symmetry of physics, it is  an important task to test  the EEP with the highest possible accuracy. This is  the main motivation of different fundamental physics experiments and in particular space experiments  
such as 
MICROSCOPE \cite{Touboul2017,TOUBOUL2001}, ACES \cite{Cacciapuoti2009}, STE-QUEST \cite{Altschul2015}, QTEST \cite{QTEST16}. 

Conceptually,
the EEP is divided into 3 sub-principles, the Universality of Free Fall (UFF), the Local Position Invariance (LPI), and the Local Lorentz Invariance (LLI), that all need to be tested independently \cite{Will2006,Altschul2015,Wolf2016}. SAGE has the potential to test all three sub-principles with unprecedented accuracy. Two of them are discussed below.

\vspace{6mm}
\subsubsection{Test of LPI through the gravitational redshift}\index{GRS:Science}

GR predicts that when comparing (e.g. using light signals) the frequencies of two 
separated 
clocks submitted to different gravitational fields  the 
observed
frequency difference will be a function of the gravitational fields at 
the locations of
the two clocks. This effect is known as the ``gravitational redshift". 

If LPI is violated there will be some anomalous coupling between the standard matter fields (the atoms in the clocks) and the ambient gravitational field, leading to a behaviour of the observed clock frequencies different from the one predicted by GR \cite{Wolf2016}. Tests of the gravitational redshift are the most prominent and precise tests of LPI at present, with 
lowest fractional 
uncertainty coming from 
clocks onboard two 
Galileo satellites on eccentric orbits: $2.5\times 10^{-5}$ \cite{Delva2018, Herrmann2018}. This results is expected to be improved by the ACES space mission to the low $10^{-6}$ level. Depending on the payload and orbital configuration, SAGE has the potential to improve on ACES by several orders of magnitude.


\subsubsection{Test of UFF and spin-gravity coupling using bosonic and fermionic Sr isotopes \label{Spin:Science}}

Since Galileo, physicists have experimentally tested the astonishing fact that all  objects fall identically, irrespective of their composition, internal structure, or mass. Such tests have continuously evolved over the centuries, not only improving their accuracy, but also diversifying the types of objects used. Best laboratory tests have relative uncertainties of $10^{-13}$ and use macroscopic test masses of different types \cite{Schlamminger2008}. Lunar laser ranging has recently reached an uncertainty of $5\times10^{-14}$~\cite{Hofmann2018}. First results from the MICROSCOPE space mission showed an improvement to the $1.3\times10^{-14}$ level~\cite{Touboul2017}. In parallel efforts are ongoing to use radically different test objects 
such as antimatter at CERN \cite{Doser2010,Perez2012}, comparisons of atoms with macroscopic test masses~\cite{Peters1999,Merlet2010,Poli2011}, atoms of different species~\cite{Zhou2015,Tarallo2014a,Schlippert2014,Bonnin2013,Fray2004}, atoms in different internal states~\cite{Rosi2017,Duan2016} and atoms in superposition states~\cite{Rosi2017}. Of particular interest in this 
context
is to perform experiments with different isotopes of Sr atoms that have different intrinsically quantum degrees of freedom, namely spin, which would bring to light fundamental spin-gravity couplings that are not detectable with classical objects \cite{Tarallo2014a}. SAGE has the potential to significantly improve on these tests, and in particular on those that use non-classical states of matter in quantum superpositions of internal states and/or with different intrinsic spins. The former require special laser pulses that can simultaneously address both internal states \cite{Roura2018}.



\subsection{Investigating quantum correlations and testing Bell inequalities for different gravitational potentials and relative velocities \label{Entangle:Science}}

A space mission gives the opportunity to test fundamental features of Quantum Mechanics in a context not accessible on ground. An experimental configuration based on two satellites is an extremely attractive scenario, 
because it avoids 
the terrestrial atmosphere and provides large relative velocities, large distances, and large gravitational potential difference. This expands the conditions in which nonlocal correlations are 
currently
observed, allowing for testing the model of instantaneous collapse of the quantum wave function and the ``spooky action at a distance" \cite{Rideout2012}.

Quantum Physics is one of the most successful contributions to science introduced in the twentieth century and, at the same time, a very  puzzling one. Among its concepts, the quantum entanglement  between two systems
is the ``characteristic trait" of quantum mechanics, according to the words of of one of the founding fathers, Erwin Schr\"odinger~\cite{schr35pro}.
A bipartite entangled state between two systems $A$ and $B$ is defined as a state that cannot be written as a product
of two states, one belonging to system $A$ and the other to system $B$. Formally, an entangled state $\ket{\Psi}$ is characterized by
the following equation:

\begin{equation}
\ket{\Psi}_{AB}\neq\ket{\phi}_A\otimes\ket{\chi}_B\ .
\end{equation}

Examples of bipartite entangled states are the so called Bell states. For polarization entanglement they can be written as
\begin{equation}
\label{BellStates}
\begin{aligned}
\ket{\Psi^\pm}&=\frac{1}{\sqrt{2}}(\ket{H}_A\otimes\ket{V}_B\pm \ket{V}_A\otimes\ket{H}_B)\ ,
\\
\ket{\Phi^\pm}&=\frac{1}{\sqrt{2}}(\ket{H}_A\otimes\ket{H}_B\pm \ket{V}_A\otimes\ket{V}_B)\ ,
\end{aligned}
\end{equation}
where $\ket{H}$ and $\ket{V}$ are the horizontal and vertical polarization states.

Entangled states are characterized by correlations that cannot be explained by classical physics. Such ``non-classical'' correlations  
lead to the violation of the so called Bell inequality~\cite{bell64phy}. 
If nature could be modeled with a ``local-hidden-variable'' model 
(a model in which some ``hidden'' variables determine the outcome of the measurement), the following inequality should be 
satisfied~\cite{clau69prl,clau74prd}
\beq
S=\langle A_1\otimes B_1\rangle + \langle A_1\otimes B_2\rangle + \langle A_2\otimes B_1\rangle - \langle A_2\otimes B_2\rangle \leq 2 \ .
\eeq
In the above equation $A_i$ and $B_i$ are dichotomic observables (i.e. with two outcomes, $-1$ and $+1$) measured
by Alice and Bob (the two subsystems) and
the term $\langle A_i\otimes B_j\rangle $ represents the correlation of the outcomes.
 
 Entangled states violate the above inequality: indeed, by using one of the four Bell states written in \eqref{BellStates}, and
 properly choosing the observables $A_i$ and $B_j$, it is possible to achieve a Bell parameter $S=2\sqrt{2}$.
In 2015, for the first time, Bell's inequality has been violated in the lab without additional assumption (the so called {\it loophole-free} violations)
\cite{hens15nat,gius15prl,shal15prl}.
 
According to standard quantum mechanics, there is no upper bound for the distance between two entangled systems
and the Bell inequality should be violated at arbitrary separation between the two subsystems $A$ and $B$.
On ground, the distance between two systems cannot be larger than a few hundred km due to optical 
losses in fibers or to the Earth's curvature for free-space propagation.
Current experiments on the ground have reached the distance of 144 km~\cite{sche10pnas}. 
 The chinese mission ``Quantum Experiments at Space Scale'' demonstrated in 2017 a 
 satellite-mediated
 distribution of entangled photon pairs between two locations separated by 1203 kilometers on Earth \cite{yin2017science}. 
Through two downlinks from the satellite Micius to two ground stations located in Delingha and Lijiang, the experimental 
violation of a Bell inequality was observed, by obtaining a Bell parameter of $S=2.37 \pm 0.09$.
With the same satellite, a space-to-ground quantum key distribution with secret key rate at kHz level was achieved \cite{liao2017nature}.

The SAGE mission configuration will allow to test quantum entanglement and thus quantum mechanics over unprecedented distances.
Any violation of the Bell inequality $S\leq 2$ will certify the presence of large-distance entanglement. 

Moreover, SAGE will address the fundamental problem of the unification of quantum mechanics and general relativity. By testing quantum correlations in a not-flat and varying gravitation background, SAGE  will indeed test Quantum Field Theory in curved spaces, the first step towards a Quantum Gravity Theory.
By using  elliptical orbits 
 that provide varying gravitational potentials, it will be possible to test alternative theories of quantum-decorrelation due to distance/time and gravitation potential. 
Moreover, since the detectors on the SAGE satellites are in relative motion, this will give insight on the wavefunction collapse. SAGE will test interpretations of the wavefunction collapse when the relative time ordering of the measurement events is not well defined~\cite{ride12cqg}.
For two spacelike separated events, the concept of simultaneity is indeed dependent 
on the reference frame.
Depending on the relative motion between the two observers each measuring one photon of the entangled pair, 
they may each measure their particle later than the other 
(the so called 'after-after' scenario), or they each measure their particle earlier than the other (‘earlier-earlier’).
These measurements will challenge the interpretation of entanglement as a non-local
influence of the ‘first’ measurement on the outcome of the ‘second’ measurement.

\subsection{Defining an ultraprecise frame of reference for Earth and Space, comparing terrestrial clocks, using clocks and links between satellites for optical VLBI in Space \label{Reference:Science:reference}}

The definition of a reference frame for Earth and Space is fundamental to geodesy and radio astronomy. Currently, VLBI from Earth is used for this purpose. SAGE can extend this possibility, by providing this information in particular to other spacecrafts.

The comparison of  optical clocks on ground at the $10^{-18}$ to $10^{-19}$ level will serve to establish a new definition of the second, to test the constancy of fundamental constants \cite{Safronova2019}, and to enable relativistic geodesy at the 0.1 cm level.  

For the application to VLBI in observational astronomy, two SAGE satellites containing optical (or infrared) telescopes are connected into a VLBI configuration. The very large baseline will enable an extremely high angular resolution.  




\section{Scientific requirements \label{Scientific requirements} }

\subsection{Observing  Gravitational Waves - Scientific requirements  \label{GWs:Requirements}}


As discussed in  section \ref{GWs:science}, a major goal of SAGE  is to observe GW signals from sources in the $10^{-3}$~Hz - 10~Hz  frequency range  with sufficiently high signal-to-noise ratio. 
%

Essential requirements are expressed in terms of spectral sensitivity to GW induced strain, and are driven by the expected signal strength. The required sensitivity in turn determines the requirements on key system parameters. 
Similarly to laser interferometric GW detectors, the proposed atomic sensors are sensitive to the strain amplitude $h$, which scales as the inverse distance $1/r$ between the source and the detector. This implies that any proportional sensitivity gain $G$ results in a corresponding improvement $G^3$ in the observed space volume.  As illustrated in section \ref{GWs:science},  a space GW detector should reach a strain sensitivity of the order of  $10^{-20}$ over two or three frequency decades in the mHz range, in order to allow the observation of a significant number of sources. 


The proposed SAGE configuration enables two possible implementations of the GW atomic detector.
In the scheme based on light-pulse atom interferometry on freely falling Sr atoms, the GW induced strain is imprinted on the differential phase on the two atom interferometers. 
In the scheme with Sr optical lattice clocks, the effect of GWs is detected as a Doppler shift in a differential frequency measurement. 

Such schemes are well suited for a wide range of inter-satellite distances; we  consider in particular the case of short baseline, i.e. $\sim10^3$\, km,  long baseline, i.e. $\sim10^6$\,km, and up to $\sim10^7$\,km for the lattice clocks scheme. 
The two proposed schemes share most of the  equipment and have similar requirements but some requirements on system parameters depend  on the 
measurement method and on the  baseline. 

The first concept for the atom-based GW antenna is to compare two light-pulse atom interferometers separated by a long baseline. To implement the atom interferometers, laser pulses are used to realise beam splitters and mirrors for the atom de Broglie waves. In a single baseline detector, the light pulses are sent from alternating directions and interact with the atoms on both ends. In this scheme, the  difference in phase between the atom interferferometers depends on the variations of the light propagation time;  by probing the phase difference,  modulations in the light travel time induced by GWs can be detected. Importantly, since the same laser pulses interact with atoms on both sides of the baseline, the common laser phase noise is substantially suppressed. This enables a detector consisting of two  spacecrafts, each with its own source of ultracold atoms.
The atom interferometers remain inside (or nearby) the local satellite, while telescopes mounted on each satellite are used to send the atom optics laser pulses across the baseline to interact with the atoms on both ends. 

The  schemes initially proposed for GW detection using atom interferometry assumed the  laser beam to be collimated, thus limiting the  baseline length $L$ to less than the Rayleigh range $z_R$ of the laser: $L \leq 2z_R =2\pi w^2/\lambda$, where $w$ is the radial beam waist and $\lambda$ the laser wavelength. A Rabi frequency $(\Omega/2\pi)\sim$~kHz  with meter-size telescopes and a laser power of 10 W leads to a baseline length $L \sim 10^3$~km. For comparison,  LISA  baseline is $ \sim 10^6$\,km.
Atom interferometry detectors  with $10^3$~km baseline can reach comparable sensitivity to LISA with a potential drastic simplification of the inter-satellite link. At the same time, such a short baseline will make some of the proposed secondary objectives feasible, as discussed in the following sections. 

On the other hand, increasing the baseline would give several advantages. Increased detection sensitivity could allow for science reach beyond LISA targets, potentially even giving access to signals of cosmological origin such as the predicted primordial gravitation waves generated by
inflation. In addition to substantially enhanced signal strength, the size of many background noise sources are suppressed. A source of local acceleration noise  $\delta a$ results in an effective strain response proportional to $\delta a/L$; therefore, a  longer baseline can reduce the  requirements needed to control several backgrounds. 
In addition, for the same  GW signal strength, a larger baseline can reduce the need for large-momentum-transfer  and other techniques for signal enhancement, leading to a simplification of the interferometer operation.
For a  configuration with  $\sim10^6$\,km baseline,  SAGE   assumes a scheme with intense lasers at each end of the baseline  for the atom interferometers. These
 independent local lasers are connected using reference lasers beams
which propagate between the two spacecraft;  the local laser sources are kept phase-locked to the incoming reference lasers. With such a  scheme, collimation  requirements for the reference beams are drastically relaxed since the phase lock can be done with an intensity much lower than
the one needed to excite the atomic transition.

In the long-baseline atom interferometry configurations described in section \ref{GW_concept}, an essential consideration is the  noise performance of the phase-lock between the local oscillator in each satellite and the reference beam. This in turn imposes a requirement on the telescope diameter $d$ for limiting the beam divergence and keep the photon shot noise below the atomic quantum noise limit \cite{hogan_atom_2015}

\begin{equation}
d=28\,\hbox{cm}\left(\frac{L}{2\cdot10^9\,\hbox{m}} \right)^{\frac{2}{5}}
\left(\frac{1\,\hbox{W}}{P_t} \right)^{\frac{1}{10}}
\left(\frac{0.2\,\hbox{Hz}/7}{f_R/n_P} \right)^{\frac{1}{5}}
\left(\frac{\bar{\delta\phi_a}}{10^{-3}/\sqrt{\hbox{Hz}}} \right)^{\frac{2}{5}}
\end{equation}
where $P_t$ is the output laser power, $f_R$ is the cycle repetition rate, $\bar{\delta\phi_a}=\sqrt{1/N_a}$ is the spectral density of atomic quantum noise and $n_P$ the number of light pulses. A conservative design assumes a $L=2\cdot10^9$\,m baseline, a laser link limited by photon shot-noise  with
a laser power of 1\,W,  telescopes diameter $d = 30$\,cm, and a
sampling rate of 0.2\,Hz. With $2\hbar k$ atom optics (corresponding to $n_P=7$) and an atomic shot-noise  
$\bar{\delta\phi_a}= 10^{-3}$\,rad/$\sqrt{\hbox{Hz}}$ the peak sensitivity reaches 
$\sim3\cdot10^{-20}/\sqrt{\hbox{Hz}}$. 
For the atom source, an ensembles of $7\cdot10^6$ atoms with  20\,pK longitudinal expansion velocity, is assumed allowing
for  a Rabi frequency of 60 Hz. Such criteria
can be met with realistic improvements of the existing technology. Large-momentum-transfer beam splitters would enhance the sensitivity in proportion to the number of photon recoils.

Other sources of noise are the jitter in timing  $\bar{\delta t_d}$ and the frequency noise $\bar{\delta\omega}$ of the pulses from the phase-locked LO laser. In particular,   the requirements for the long term frequency stability  of the LO laser can be relaxed if the pulses from the LO are  synchronised with the incoming reference pulses, keeping the delay $t_d$, introduced by the phase lock of the two local oscillator lasers, small  \cite{hogan_atom_2015}. In practice,
$t_d\sim10$\,ns with an RMS noise $\delta t_d\sim1$\,ns appears feasible, showing that the LO pulse timing constraints can be fulfilled.

Noise can also be introduced by satellite and laser beam pointing jitter; however, typical requirements ($\sim10^{-6}$\,rad/$\sqrt{\hbox{Hz}}$) are not particularly stringent.

In the differential frequency measurement scheme with two optical clocks sharing the same clock laser, a minimum detectable fractional frequency difference of $\sim1\cdot10^{-20}/\sqrt{\rm{Hz}}$ is assumed \cite{Kolkowitz2016}. Atoms are confined in an optical lattice trap, posing stringent requirements on the position stability of the retroreflecting mirror employed to generate the trap standing wave; it turns out that the mirror must be drag-free with residual motion at the level of the LISA test masses. The AI approach instead allows to relax the requirements for drag-free control.
The AI measurement requires, on the other hand, low expansion velocities for the atomic sample, of the order of 20 pK. Comparable collimation performance is already achieved using delta-kick collimation techniques \cite{Chu1986pro,AmmanPRL1997,Muentinga13PRL,KovachyPRL2015}. A detailed study of the atomic sources requirements in the case of an AI operation is reported in \cite{Loriani2019}. In the optical clocks scheme the atoms can be loaded into the optical lattice trap at microkelvin temperatures.


While the requirements on the SAGE configuration and instruments are driven by the primary goal of GW observation, the other important experiments proposed as secondary objectives can be performed with a minor impact on the requirements. In order to optimize the performances for the different experiments, a mission in two  phases with different orbits can be considered.

\subsection{Searching for Dark Matter  - Scientific requirements \label{DM:Requirements}}
Clumpy DM can be formed from macroscopic constituents, such as Q-balls~\cite{KusSte01}, solitons, or topological defects (monopoles, domain walls, or strings). Another possibility are DM ``blobs'': particles with long-range Yukawa-type interactions that couple feebly to standard model particles and fields~\cite{Grabowska2018}. Such DM ``blobs'' would be perceived as spherically-symmetric monopole- or Q-ball-like objects.  For concreteness, below we focus on topological defects. 

The large size of the SAGE space mission provides a unique opportunity to search for spatially-extended DM objects, such as the so-called topological defects (TD), which cannot be detected by most currently ongoing and future planned DM searches~\cite{Derevianko2014}. Indeed, the existence of hypothesized, beyond-standard-model self-interacting cosmic fields, which we shall denote as $\phi$, may lead to the formation of spatially extended (or ``clumpy'') DM objects. The stability of such objects (monopoles, strings and domain walls) is assured  by the topological arguments~\cite{Vilenkin1985}. TDs were searched for via their gravitational effects, including gravitational lensing~\cite{Vilenkin1994,Schneider1999,Cline2001}. Constraints on the TD contribution from the fluctuations in the cosmic microwave background  were set by Planck~\cite{Ade2014a} and BICEP2~\cite {Ade2014,Moss2014}. So far, the existence of TDs cannot be confirmed nor ruled out. While the exact nature of TDs depends on the details of specific models, in general the spatial extent $d$ of the DM object is related to the Compton wavelength of the particles constituting the DM field: $d\sim \hbar/(m_\varphi c)$, where $\hbar$ is the reduced Plank constant and $m_\varphi$ is the mass of the particle.

For the  search of DM in the form of  monopoles, according to the model in~\cite{Derevianko2014}, a network of atomic clocks must be  dense and large in order to maximize the DM-network encounter rate and cross-section,  and in order to detect the smallest DM clumps. For the monopoles, the distance between atomic clocks on the two SAGE satellites sets the smallest size to which the network can be sensitive.   For domain walls this consideration does not apply as the wall will sweep through both satellites. Walls can also contribute to dark energy due to the dark energy equation of state. If the walls form close on themselves forming cosmic bubbles, their equation of state is that of the pressureless cosmological fluid, i.e., that of the dark matter~\cite{Roberts2017-GPS-DM}).

\begin{figure}[h]
\centering
	\includegraphics[width=0.5\textwidth]{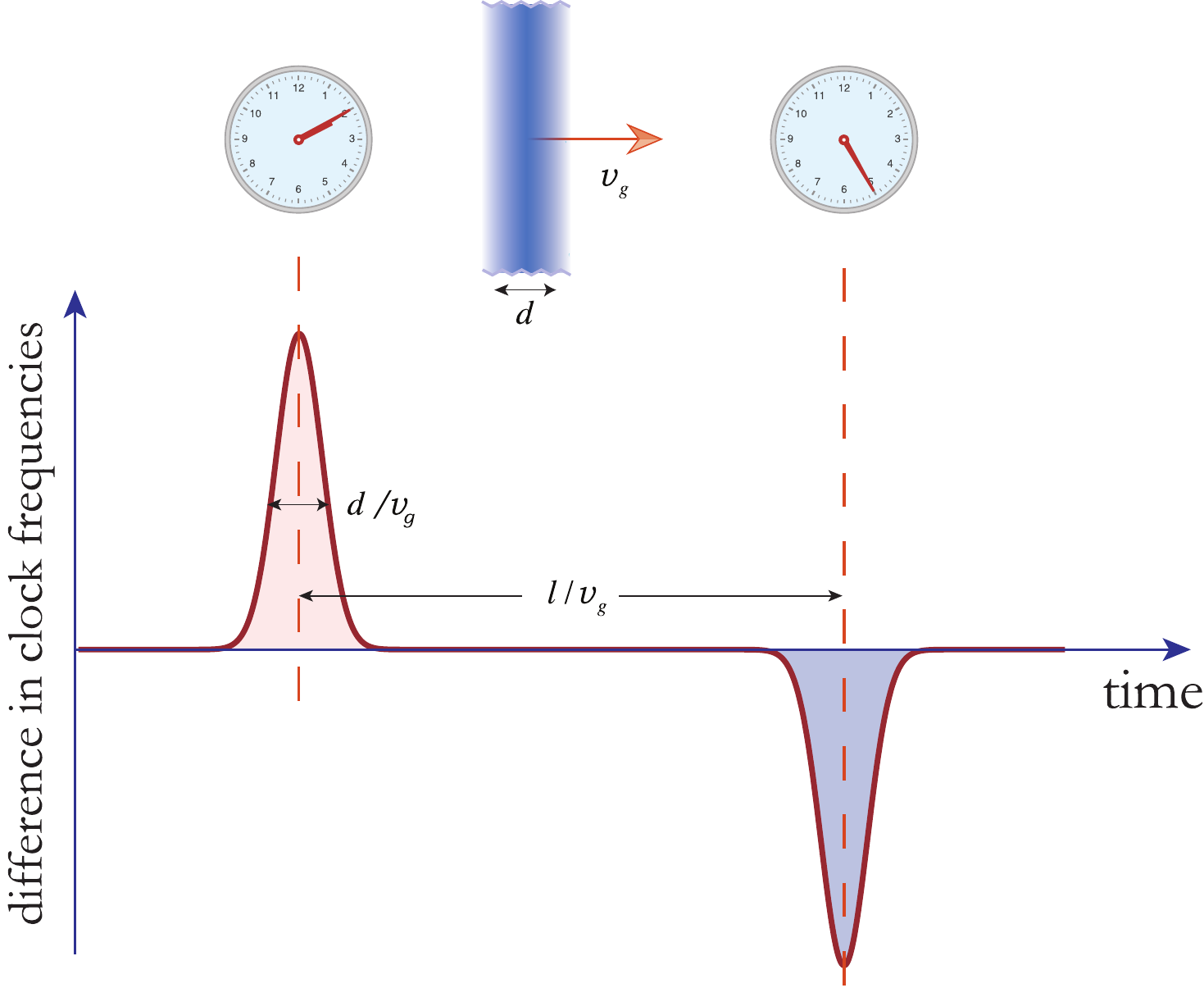}
	\caption{Time-dependence of the  frequency difference  between two identical clocks induced  by a  domain-wall (of thickness $d$) sweep at galactic velocities $v_g \sim 300 \, \mathrm{km/s}$.
	The clocks are separated by a distance $l>d$. Each DM-induced spike lasts for $\sim d/v_g$ and the second spike is delayed by $l/v_g$. Figure taken from~\cite{Roberts2017-GPS-DM} (http://creativecommons.org/licenses/by/4.0/).}
\label{Fig:twoclocks}
\end{figure}

According to the standard halo model,  the velocity distribution of DM objects in the galactic rest frame  is quasi-Maxwellian, with virial velocity  dispersion $v\simeq 270\, {\rm km/s}$ and a cutoff above the galactic escape velocity  $v_{\rm esc} \simeq 650\, {\rm km/s}$ (see, e.g., Ref.~\cite{Freese2013}). 
This distribution is modified by an addition of the Sun's motion relative to the galactic  center, about $230\, {\rm km \,s^{-1}}$. 
Therefore, we can consider a ``wind'' of topological defects impinging on the mission satellites, with typical galactic velocities $v_g  \approx 300 \, \mathrm{km/s}$ coming from the direction of Cygnus constellation.  
A  DM signal can be expected as a  propagation of clock ``glitches'' at galactic velocities through the SAGE mission satellites. 
The powerful advantage of  the distributed clock configuration is that the clock perturbations which are not due to DM in general do not show this signature. The  solar wind is an  effect that has a propagation velocity similar to $v_g$  but it can be discriminated due to its  directionality.

The relevant  non-gravitational interactions between the clock atoms and  DM  can be parameterized in terms of  shifts in the  values of fundamental constants \cite{Derevianko2014}. For example, 
\begin{equation}
\alpha \to \alpha'(\mathbf{r},t) = \alpha\left(1+\frac{\delta\alpha_{\rm DM}(\mathbf{r},t)}{\alpha}\right),
\end{equation}
where $\alpha = e^2/\hbar c \approx1/137$ is the  electromagnetic fine-structure constant.
In the assumption of quadratic coupling to DM \cite{Derevianko2014}, that is relevant to the fields of either $Z_2$ or $U(1)$ internal symmetry,
\begin{equation}
\frac{\delta \alpha (\mathbf{r},t)}{\alpha} = \frac{\phi(\mathbf{r},t)^2}{\Lambda_\alpha^2},
\label{Eq:X-shift}
\end{equation} 
to first order in $\phi^2$, where $\Lambda_\alpha$
is the effective energy scale for the DM objects that determines the strength of the coupling of DM to standard model (SM) particles.
The fine-structure constant is modified only when the core of the DM object interacts with the atomic clock.
This transient shift in the effective fundamental constant value leads  to a shift in the atomic  frequency referenced by the clocks that can be expressed as
\begin{equation}
\frac{\delta \omega (\mathbf{r},t)}{\omega_c} =  K_\alpha\frac{\delta \alpha (\mathbf{r},t)}{\alpha}, \label{Eq:ClockFreqPert}
\end{equation} 
where $\omega_c$ is the unperturbed  frequency of the clock
and $K_\alpha\approx 2$ is a coefficient quantifying the Sr clock transition sensitivity  to changes of $\alpha$.

From Eqs.~(\ref{Eq:X-shift},\ref{Eq:ClockFreqPert}), the maximum frequency excursion, $|\delta {\omega _{\max }}|$,  is related to the amplitude $A$  of the field inside the defect as
$|\delta {\omega |_{\max }} = K_\alpha {\omega _c}{A^2}/{\Lambda_\alpha^2}$ with the  energy scale $\Lambda_\alpha$. 
In the assumption that a specific TD  saturates the known DM energy density $\rho_\mathrm{DM}$, 
$A^{2}=\hbar c \rho_\mathrm{DM} \mathcal{T} v_{g} d$ ~\cite{Derevianko2014}, where $\mathcal{T}$ represents the time between consecutive interactions of the atomic clock with DM objects. If none of the frequency excursions is DM-induced, one may set a limit on the energy scale 
\begin{equation}
\Lambda _\alpha^2 > \hbar c K_\alpha {\rho _\mathrm{DM}}  {v_g}\, \frac {\omega _c}{\left| {\delta \omega } \right|_{\max }} \mathcal{T}_\mathrm{obs} d
, \label{Eq:Limit}
\end{equation}
where $\mathcal{T}_\mathrm{obs} $ is the total observation time (mission duration).
In order for  Eq.~(\ref{Eq:Limit}) to be valid,  the possibility of discriminating between the DM induced frequency changes and the  clock noise is crucial. This can be achieved by measuring the time delays between DM events at the nodes of the network. If we consider two spatially-separated atomic clocks (Fig.~\ref{Fig:twoclocks}), 
a frequency shift induced by DM (Eq.~(\ref{Eq:ClockFreqPert})) will produce a distinct pattern. 
The velocity of the sweep translates into the time delay between  DM-induced spikes; that must be consistent with the boundaries from the standard halo model.

We close this subsection with a brief discussion of ultralight DM candidates based on nearly homogeneous oscillating scalar fields, which lead to small-amplitude oscillations of the fundamental constants at the Compton frequency of the scalar field.
A comparison of clocks of different nature (for example, optical vs microwave) is needed for detection schemes involving the comparison of co-located clocks. In contrast, the same atomic species can be employed when considering atom interferometers separated by long baselines \cite{Arvanitaki2016}. In this respect, the sensitivity planned for the SAGE atom interferometers will allow an improvement by several orders of magnitude in the investigation of possible effects of such ultralight DM candidates.

\subsection{Testing General Relativity  - Scientific requirements}

The LPI test is best performed in the optical lattice clock scheme, with the two clocks compared using the laser signals that are also used for GW detection. The envisioned clock performance expressed as fractional frequency fluctuations is assumed to be as in \cite{Kolkowitz2016} of order $10^{-20}/\sqrt{\mathrm{Hz}}$ down to $10^{-3}$~Hz. To take advantage of the clock stability a modulation of the gravitational field at the clock location is required.

A first experiment could consist of having one clock already on its cruise phase and the other still in a GTO Earth orbit. For definitiveness we  consider an orbit as in the STE-QUEST M4 proposal (apogee at 33600 km, perigee at 2500 km). Comparing the two clocks over half an orbital period of the GTO orbit (5 hours) yields a test of the gravitational redshift in the Earth field at potentially $10^{-13}$, a 7 orders of magnitude improvement with respect to ACES. However, the experiment will be limited by the orbit determination uncertainty in GTO orbit. With centimetric orbit determination one could still improve the expected ACES result by 3 orders of magnitude or more. 

A similar, but complementary, experiment could be carried out in the field of the sun \cite{Wolf2016} when the two satellites are in sun orbit, provided that those orbits are slightly eccentric. Note that even a modest eccentricity (say 0.01) leads to a potential test at the $10^{-12}$ level, 10 orders of magnitude better than the best present direct tests and 6 orders of magnitude better than the best present ``null" tests. Again, the final limit is likely to come from the knowledge of the satellite-to-sun distance. An uncertainty of 1 m for the satellite-to-sun distance would lead to a limit two to three orders of magnitude less stringent but still of strong interest.

%

\subsection{Investigating quantum correlations and testing Bell inequalities for different gravitational
potentials and relative velocities - Scientific requirements}\index{Entangle:Requirements}

For the observation of quantum correlations and test of the violation of the Bell inequality, a high purity and high brightness source of entangled photons  is needed. This source is based on a nonlinear crystal and a high coherence and high power laser, that may be shared from the GW experiment.
The entangled photon to be transmitted will be generated at the same wavelength that is optimized for the GW experiment.
Detection of single photons will be realized with high efficiency and low noise single photon detectors combined with the polarization analyzer \cite{Vallone2015}. The detection time will be referenced to the ultraprecise atomic clocks used for the GW experiment.
It is worth noting that these interesting experiments could be realized with a minor impact on the main mission configuration and that  the additional required components have already a high technology readiness level.

\subsection{Defining an ultraprecise frame of reference for Earth and Space, comparing terrestrial clocks,
using clocks and links between satellites for optical VLBI in Space - Scientific requirements}\index{Reference:Requirements}

One or two SAGE satellites will contain two optical links and a frequency comb, so that two distant ground clocks can be compared with the on-board clock and with each other in common view using two-way frequency-comb based time transfer.
The SAGE satellites will preferably be in an orbit with small velocity compared to ground, thereby minimizing the 2nd order Doppler shift correction \cite{pet05}. The satellites will also preferably be in a high orbit, so that ground clocks located at intercontinental distances can be compared, and furthermore the common-view duration is long. An orbit close to geostationary appears as a good choice for this application.

%




\section{\label{Measurements_concept} Measurement concept}

\subsection{\label{GW_concept}  Observing Gravitational Waves -  Measurement concepts}\index{GWs:Measurements}

The proposed detector is based on a long baseline multi-satellite link between ultracold atomic Sr systems. The basic configuration consists of two satellites, and the link is performed with laser radiation inducing the 698 nm clock transition in atomic strontium. 

We consider two  approaches to provide sensitivity to the GW induced strain: In the first method, based on momentum transfer from the optical field to the atomic sample, Sr atoms act as test masses in an atom interferometry scheme. In the second approach, Sr optical atomic clocks provide ultra-stable frequency references and the GW signal is detected from the induced Doppler shift by synchronized two-clock comparison.

\subsubsection{\label{AI_concept} Light-pulse atom interferometry}\index{GWs:AI_Measurements}

\begin{figure}[tt]
\centering
\mbox{%
\begin{minipage}{.60\textwidth}
\includegraphics[angle=0, width=\textwidth]{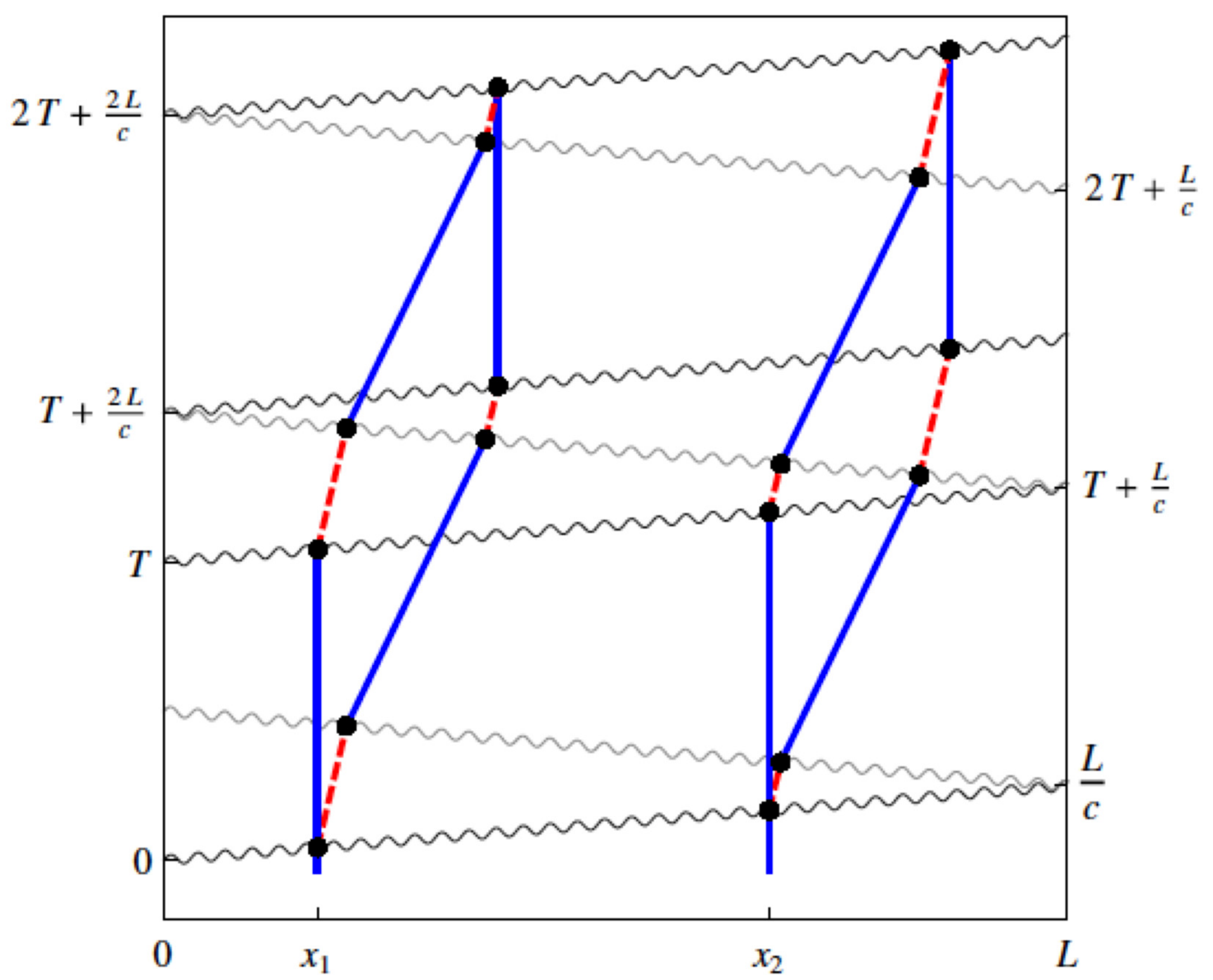}
\end{minipage}%
\quad
\begin{minipage}[c]{.39\textwidth}
\caption{ 
Space-time diagram of the trajectories for two atom interferometers based on single-photon transitions between the ground atomic state (solid line) and the excited state (dashed line). Laser pulses (wavy lines) traveling from alternating sides of the baseline are used to split, redirect, and recombine the atom de Broglie waves, thus producing atomic interference signals that are sensitive to the modulation of the light travel time caused by gravitational waves. (From \cite{Graham2013a}.)
} \label{fig:gwsetup}
\end{minipage}%
\quad
}
\end{figure}

%
%

A scheme of the atom interferometry GW antenna is shown  in Fig.~\ref{fig:gwsetup}. It is similar to the the one used for atom  gravity gradiometers  \cite{SnaddenGG,Sorrentino2014,Chiow2016}. Clouds of cold atoms at both ends of the baseline play the role of  inertial test masses with laser light propagating between the atomic clouds.  To implement atom interferometry, the light from the two lasers  are  pulsed several times for each measurement cycle.  The light pulse paths are shown as wavy lines in Fig.~\ref{fig:gwsetup}.  The lasers are separated by a large distance $L$ and the atom interferometers, represented by the two diamond-shaped loops, are operated near them.    A laser pulse transfers a momentum $\hbar k$ to the atom and switches the internal atomic state between the ground and the excited state.  Therefore, the light pulses act as beam splitters and mirrors for the atomic de Broglie waves, splitting them into a quantum superposition of two paths and then recombining them. As for an atomic clock, each atom interferometer records a  phase shift that depends on the time spent in the excited state that  is  connected to the light travel time ($L/c$) across the baseline. GWs can therefore be detected because of the modulation of the  travel time of the light.

While a single interferometer of the type described above (e.g., the interferometer starting at position $x_1$ in Fig.~\ref{fig:gwsetup}) will be affected by the laser noise, this effect is substantially suppressed by the differential measurement with the two interferometers  \cite{Graham2013a}.  The two  separated atom interferometers are realized using common laser beams and the differential phase shift is measured.  Noticeably, for each laser interaction, the same laser beam drives both interferometers. For both interferometers, the laser pulse from the first laser triggers at time $t=0$  the initial beam splitter while the pulse from the second laser completes this beam splitter at time $t=L/c$. The differential phase shift for the two interferometers contains the gravitational wave signal which is proportional to the distance between them.
The differential signal is instead virtually immune to laser frequency noise because the same laser pulses operate both interferometers.

The cancellation of laser noise is enabled by the use of single-photon atomic transitions for the atom optics, as originally proposed in \cite{Yu2011}.  Critically, each single-photon transition is driven by a single laser pulse originating from one of the laser sources.  The laser frequency noise from each pulse is then common to both atom interferometers and cancels in the differential measurement,  according to the relativistic formulation of atom interferometry  \cite{Dimopoulos2008a};  the laser phase of a pulse is set when the pulse is emitted and does not change as it propagates along the null geodesic connecting the laser to the atoms.  The proposed laser excitation scheme is based only on single-photon atomic transitions in order to take advantage of this noise immunity.  In an optical interferometer, the relative phases of the interfering optical fields serve as proxies for the  light propagation time   along the interferometer arms.  Here, instead gravitational waves are  sensed by a direct measurement of the time intervals between optical pulses, as registered  by the high stability oscillators represented by the atomic transitions.

As mentioned above, initial proposals considered collimated laser beams driving the atomic transitions, thus constraining the  baseline length $L$ to be smaller than the laser Rayleigh range $z_R$.  
Assuming  Rabi frequencies $\sim \text{kHz}$ with  telescope diameter $\sim 1~\text{m}$ and  laser power $\sim 10~\text{W}$, sets a  limit  $L \sim 10^3~\text{km}$ for the  length of the baseline \cite{Dimopoulos2008,Graham2013a}.  
Here we consider a concept for an atom interferometric GW detector that can support substantially longer baselines without requiring proportionally larger telescopes or increased laser power, following the original proposal in \cite{Yu2011conf,CHIOW2015}.
In the proposed implementation, intense local lasers realize the atom interferometers at the two ends of the baseline.  To connect the two independent local lasers, reference laser beams are used  between the two spacecrafts; the phases of the local lasers are locked/monitored with respect to the incoming wavefronts of the reference lasers. This scheme relaxes the requirement for the collimation of the reference beams because the phase measurement is done with lower intensity than the one needed to excite the atomic transition.  
Therefore, with  a modest size of the telescopes  and power of the reference beams, a baseline length similar to the one of LISA can be considered.  Since the local lasers track the phase noise of the  reference lasers, this scheme keeps the cancellation of the common mode laser phase noise  between the two interferometers thus enabling the single baseline operation.  Since the requirement  for  the phase noise rejection is effectively decoupled  from the intensity requirement,   the baseline and the atomic transition rate can be independently optimized.

\begin{figure}
    \centering
    \includegraphics[width=\columnwidth]{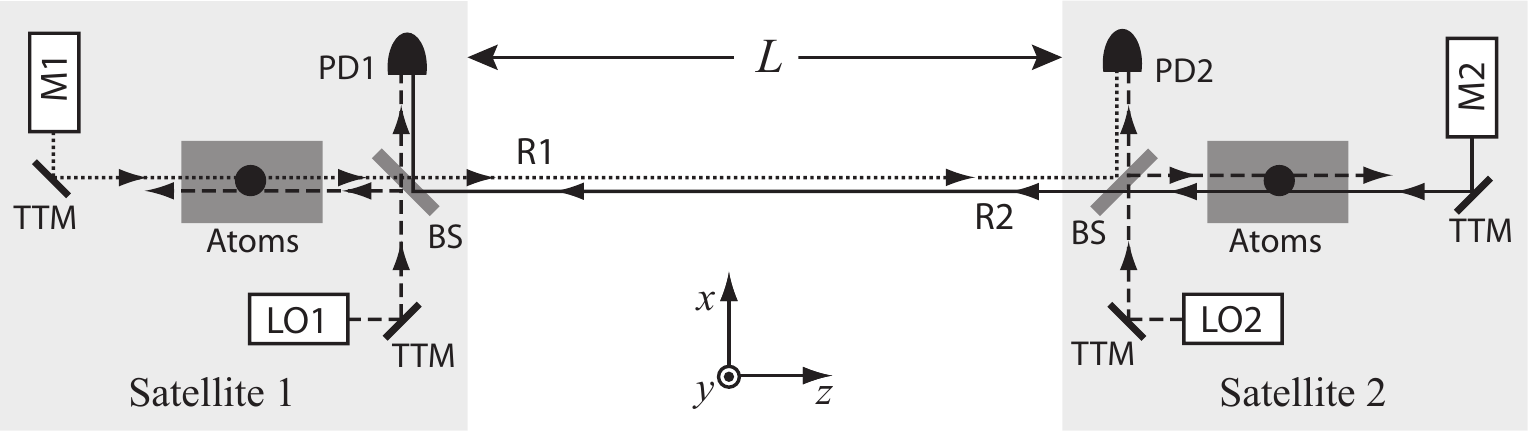}
    \caption{Proposed scheme.  The master lasers M1 and M2 generate beams which are shown as dotted and solid lines, respectively. The reference laser beams  between the satellites are  R1 (dotted line) and R2 (solid line).   The local oscillator lasers LO1 and LO2 (dashed  lines)  are phase-locked to the  reference laser beams R2 and R1, respectively. PD1 and PD2 are the photodetectors used to measure the heterodyne beatnote between the incoming reference beams R2 and R1 and the local oscillator lasers LO1 and LO2, respectively; this provides the feedback for the laser link. BS: nonpolarizing beam-splitter. TTM: Tip-tilt mirror used for controlling the pointing direction of the laser beams. Overlapping laser beams are shown as parallel beams with a small offset for the sake of clarity.
     (From \cite{hogan_atom_2015}.)}
    \label{Fig:apparatus}
\end{figure}

A conceptual schematic is shown in Fig.~\ref{Fig:apparatus}.  While this figure summarizes the basic idea, there are different promising implementation strategies, one of which is detailed 
below. An atom interferometer on each satellite  is realized with laser pulses propagating along  the positive and negative $z$ directions.  An intense master laser (M1 and M2) on each satellite  drives the atomic transitions in the local atom interferometer.  After the interaction with the atoms, each master laser beam is transmitted by the beam splitter, exits the satellite  and  propagates towards the opposite satellite.  R1 and R2 are the  beams  from satellite 1 and 2, respectively, playing the role of reference beams. 
The reference beams reaching the opposite satellite  do not need to be collimated;  in the case of very long baselines the received reference beam intensity will be too low to directly drive the atomic transition but the LO lasers driving the transition (LO1 and LO2) can be phase-locked to the incoming reference beams.  A photodetector and a quadrant detector or a camera can be used to measure the phase difference between the two beams and to characterize the spatial interference pattern.  In this way, the pointing direction  and the spatial mode of the two lasers can be matched.


\subsubsection{\label{clock_concept} Differential frequency measurement with optical lattice clocks}\index{GWs:clock_Measurements}

In the second approach, the ultracold atomic samples are employed to provide ultra-stable frequency references (optical atomic clocks), and the GW signal is detected from the induced Doppler shift by synchronised two-clock comparison.
This method relies on the most recent advances in the high resolution spectroscopy of forbidden optical transition that is used in modern optical lattice clocks.

While in atom interferometers the atomic motional degrees of freedom  measure the laser frequency changes, in this second scheme the internal states of the atomic sample (coupled by the optical clock transition and isolated from the external degrees of freedom) are used as a frequency reference to compare the frequency of the laser shared between the two spacecrafts.

\begin{figure}
	\centering
	\includegraphics[width=0.9\columnwidth]{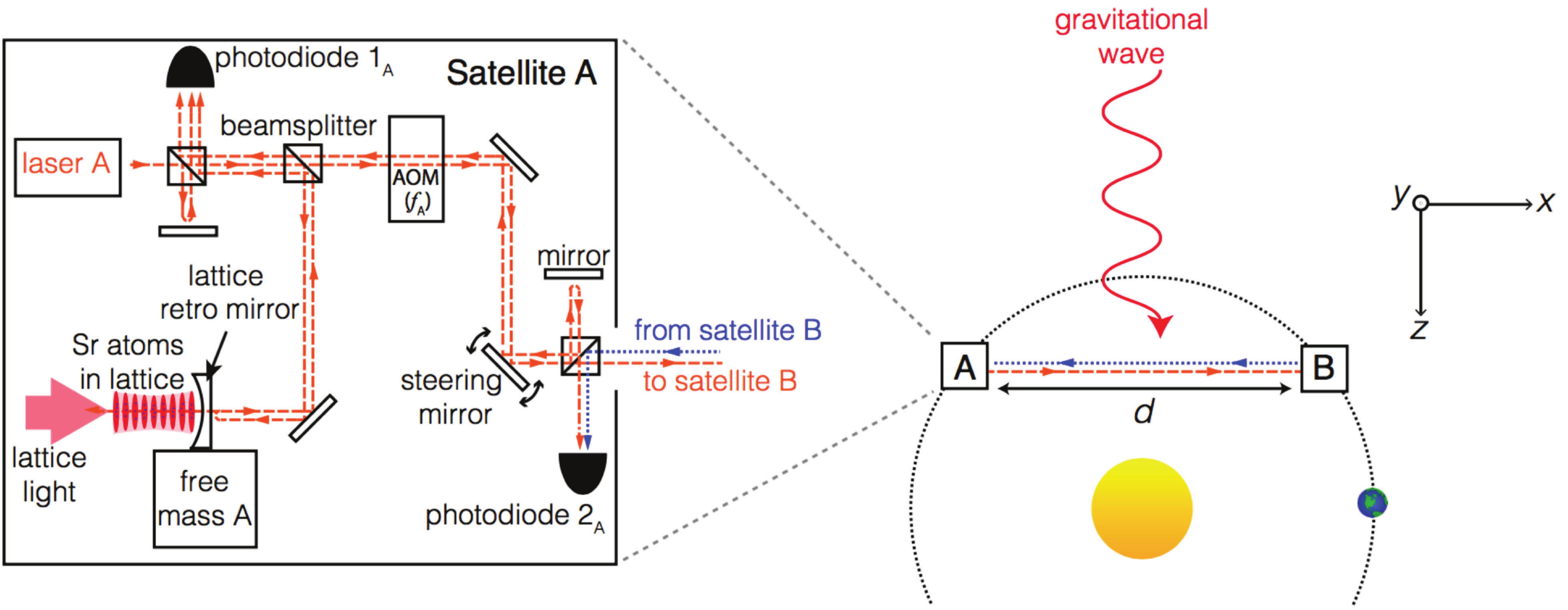}
	\caption{GW detector based on optical lattice clocks. Two  drag-free satellites (A and B) are separated by a distance $d$ in a heliocentric orbit and connected with an optical laser link. Each satellite carries on-board a free-floating test mass, an ultrastable clock laser and an optical  clock.  A mirror mounted on the  test mass defines the laser standing wave that forms the optical lattice and confines the atoms. A fraction of the laser light from satellite A (dashed line) is sent to satellite B. A set of acousto-optical modulators (AOM) and photodiodes on each satellite are used to phase lock the laser B to the laser A by heterodyne detection.   Thermal drifts and vibrations  of the optics on the satellites can  be corrected locally by feedback on the beat notes observed on the two phododiodes. A GW propagating along the z-axis, which mimics the effect of relative motion between the two free masses, can be detected by clock comparison. (From \cite{Kolkowitz2016}.) }
	\label{Fig:ye1}
\end{figure}

A schematic view of the proposed detector based on optical lattice clocks is presented in Fig.~\ref{Fig:ye1}. The detector is composed of two identical drag-free satellites, each one carrying a Sr optical lattice clock as a frequency reference. A laser link  between the two satellites  is used to phase lock the two local oscillators. 
In this way, only one laser is used to probe the two distant clocks synchronously thus reducing the sensitivity to laser phase noise. 

A + polarized GW with strain amplitude $h$ and frequency  $f_{GW}$ that propagates along the $z$-axis perpendicular to the laser link between the two satellites  mimics the effect of relative motion between the A and B  free masses; therefore, a  Doppler shift of the reference clock laser light propagating from the satellite A to B can be measured.  The atoms in the B satellite will interact with a local oscillator of a different optical frequency than the atoms in the A satellite and will accumulate a different phase. A comparison of the two clocks will show different  ticking  rates; the  fractional frequency difference between the two clocks is given by
\begin{equation}
s=\frac{\delta\nu}{\nu}= h |\sin (\pi f_{GW} d/c)|,
\end{equation}
where $c$ is the speed of light.   The optimal clock spacing is $d=\lambda_{GW}/2$,  where $\lambda_{GW} = c/f_{GW}$ is the GW wavelength, and in this case $s = h$. 

\begin{figure}[hh]
\centering
\mbox{%
\begin{minipage}{.50\textwidth}
\includegraphics[angle=0, width=\textwidth]{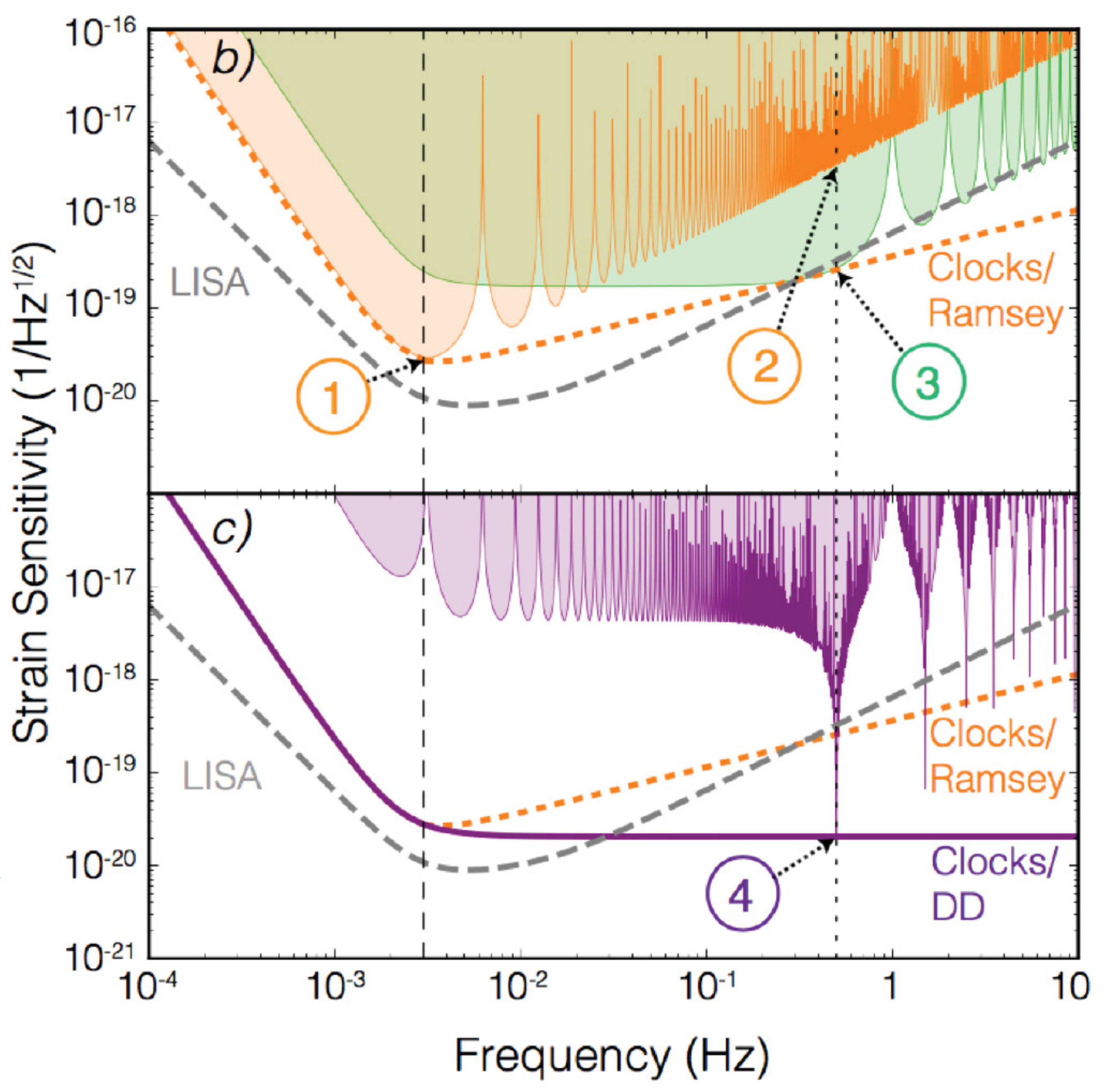}
\end{minipage}%
\quad
\begin{minipage}[c]{.50\textwidth}
\caption{ (top)  Shot-noise-limited  sensitivity of the optical atomic clock detector to a monochromatic GW using interrogating pulse sequence with two different total times, optimized for different Fourier frequencies in the mHz-Hz range.  For  sequences optimized for each GW frequency, the sensitivity envelope is also shown (dotted line) and compared with LISA. %
We consider an optical atomic clock on each satellite operated with $7 \cdot 10^6$ atoms, corresponding to a minimum detectable fractional frequency difference of $\sim1\cdot10^{-20}/\sqrt{\rm{Hz}}$ \cite{Kolkowitz2016}, and a baseline length $d=5 \cdot 10^{10}$ m optimized for $f_{GW}>3$ mHz.
(bottom) Shot-noise-limited   sensitivity of an optical clock GW detector  using a modified interrogating sequence (filled region), and the corresponding sensitivity envelope for  sequences optimized for each GW frequency (thick line). (From \cite{Kolkowitz2016}.)
 } \label{Fig:ye2}
\end{minipage}%
\quad
}
\end{figure}

A comparison of the sensitivity limited by shot-noise   for the optical atomic clock GW detector with the sensitivity of the LISA detector is shown in Fig.~\ref{Fig:ye2}.  While  LISA  will provide broadband sensitivity,  an atomic clock GW detector can perform a narrowband measurement at frequencies selected by  a dynamical decoupling  scheme; this
consists of a Ramsey sequence  with a succession of periodically spaced pulses matching  the GW frequency. 

This scheme is complementary to the one of LISA;  an optical clock GW detector might indeed be integrated with LISA and operated in parallel to it without affecting the sensitivity of either detector while increasing the observation potential. 
%
%
%
%
For example, using such a hybrid configuration, after the detection by LISA of an on-going binary inspiral at mHz frequencies, the atomic clock GW detector would enable the observation even when the GW frequency is out of the LISA detection bandwidth up to the final moments of the merger and until it can be observed by terrestrial GW detectors. 
In addition, LISA could greatly benefit from optical lattice atomic clocks onboard since the ultrastable lasers locked to the atomic reference would be an ideal LO for the optical interferometer.

With respect to atom interferometer (AI) detector, there is an advantage in terms of the required atomic temperature.
In atom interferometers, indeed,  the atoms themselves act as free-falling test mass, thus relaxing the requirements for drag-free satellite, but the atoms need to be cooled down to picoKelvin temperatures.
For the clock detector, instead, microKelvin temperatures are sufficient, greatly  simplifying the atomic cooling stage.
Finally, a laser power of about 1 W is required for the  optical link among the two satellites.

\subsection{Searching for Dark Matter  - Measurement concepts}\index{DM:Measurements}
The search of dark matter can be carried out by proper analysis of the output data from the SAGE ensemble of atomic clocks and interferometers.
The two different approaches presented in section \ref{GW_concept} are compatible with DM search methods. In particular, 
the use of atomic clocks for DM detection is discussed in \cite{Derevianko2014,Arvanitaki2015,Derevianko2016}, while the use of atom interferometry based GW detectors in space to DM search is discussed in \cite{Geraci2016,Arvanitaki2016}.

According to the model~\cite{Derevianko2014} when the atomic clock passes through the dark matter clumps, the difference of fundamental constants, such as the fine-structure constant or electron mass, inside and outside the clumps makes the clock to slow down or speed up. On the other hand, according to the model in \cite{Arvanitaki2015}, nearly homogeneous ultralight fields can lead to oscillating values for physical constants at the relevant Compton frequency. By Fourier-transforming the data from clock frequency measurements, one could detect peaks in the power spectrum thus identifying DM. The spectral profile is expected to have a characteristic lineshape, providing a unique DM signature~\cite{Derevianko2016}. In atom interferometers, time-varying phase signals from oscillatory, or dilaton-like, DM can be detected due to changes in the atom rest mass  during the light-pulses interferometer sequence \cite{Geraci2016,Arvanitaki2016}. It is expected that several orders of magnitude of unexplored parameter space for light DM fields can be probed with SAGE.

As an example, below we consider the discovery reach in the search for domain walls (or quantum bubbles, when domain walls close on themselves). This analysis follows Ref.~\cite{Roberts2017-GPS-DM}. The SAGE mission has two Sr optical lattice clocks separated by the distance $l$. As such it is only sensitive to topological defects in the form of domain walls. From Fig.~\ref{Fig:twoclocks} it is apparent that the signature is not washed out as long as the wall thickness $d < l/2$. On the lower end, because of the finite minimal clock interrogation time $\tau_\mathrm{min} \sim 1\, \mathrm{s}$ and servo-loop delays, the minimum wall thickness the SAGE mission is sensitive to is $d \sim  v_g   \tau_\mathrm{min}\approx 300 \, \mathrm{km}$. 
The intrinsic frequency noise of current clocks, characterized by the Allan deviation $\sigma_y(\tau)$, provides a conservative estimate for $\left| {\delta \omega } \right|_{\max }$ sensitivity in Eq.~\ref{Eq:Limit}.
\begin{equation}
\frac{\left| \delta \omega  \right|_{\max }} {\omega _c} <  \sigma_y(\tau) = 10^{-16}/\sqrt{\tau/\mathrm{s}} \, , \label{Eq:Allan}
\end{equation}
where $\tau$ is the clock interrogation time limited from below by $\tau_\mathrm{min}$ and from above by $\tau_\mathrm{max} = d/v_g$.
This is a conservative estimate as many noise-induced jumps will be dismissed  because they would not fit the sought signature of Fig.~\ref{Fig:twoclocks}. More sophisticated Bayesian statistics data analysis approach applicable to the SAGE mission can be found in Ref.~\cite{Roberts2018a}.

\begin{figure}[h]
\centering
	\includegraphics[width=0.75\textwidth]{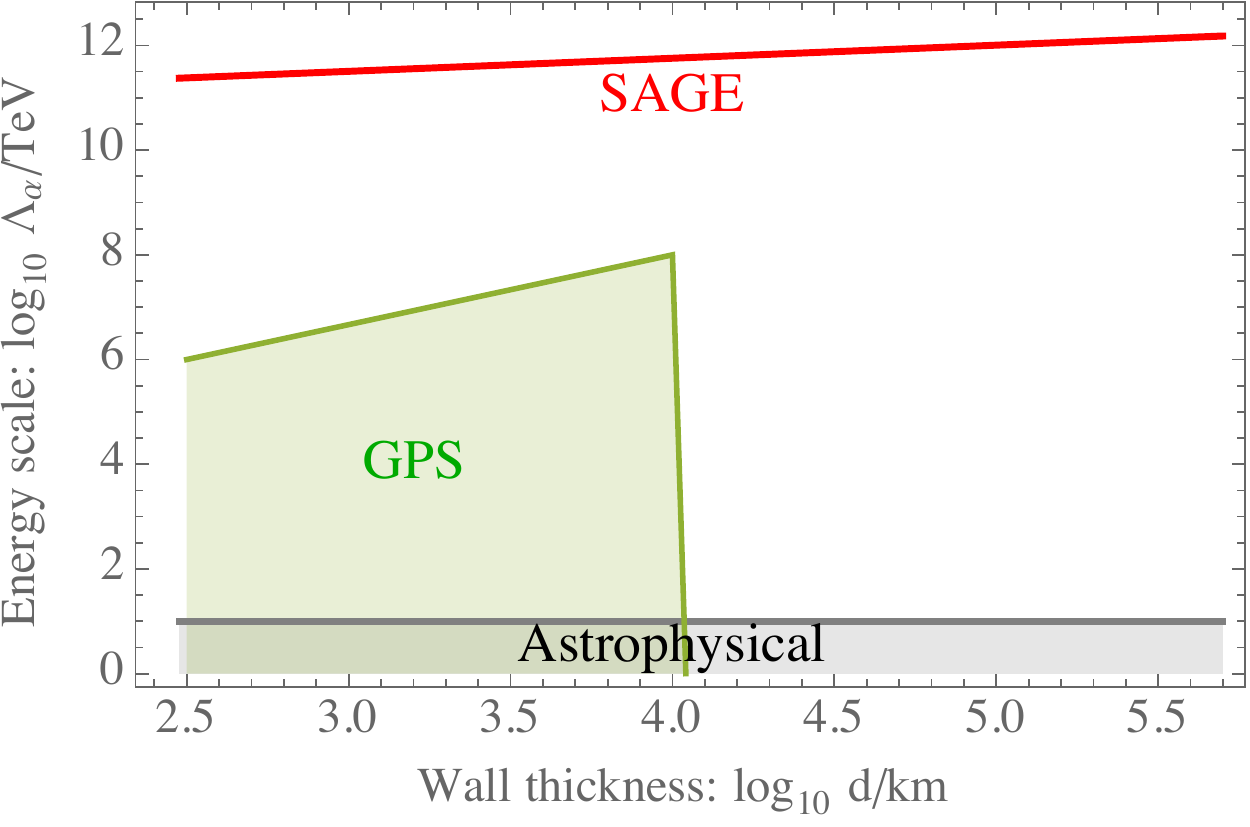} 
	\caption{Sensitivity to topological DM as domain walls for the $10^6$ km SAGE mission configuration assuming a 5-year mission duration. 
	 Current constraints  on the energy scale $\Lambda _\alpha$  of dark-matter induced variation of $\alpha$ come from Rb clocks on board GPS satellites~\cite{Roberts2017-GPS-DM} (green shaded region) and astrophysical bounds from supernova emission~\cite{Olive:2007aj} (grey shaded area). 
	The minimum wall thickness SAGE  is sensitive to is 300 km and the maximum wall thickness is $5\times 10^5$ km.
	 The wall widths can be converted into the underlying field masses $m_\varphi$ via the Compton relation $d\sim \hbar/(m_\varphi c)$.}
\label{Fig:SAGE-sensitivity}
\end{figure}

The final result for the conservative estimate of sensitivity of the SAGE mission to DM interaction energy scale is
\[
\Lambda _\alpha > 
3 \times 10^{11} \, \mathrm{TeV}\, \left( \frac{d}{10^3\, \mathrm{km}}  \right)^{1/4} \left( \frac{l}{10^6 \,\mathrm{km}} \right)^{1/2} \, ,
\]
where $d$ is the domain wall thickness and $l$ is the distance between satellites. Here we have considered a 5-year mission duration and the Allan deviation estimate~(\ref{Eq:Allan}).  The sensitivity plot is shown in Fig.~\ref{Fig:SAGE-sensitivity}. Apparently, the  SAGE mission can either lead to the discovery of dark matter in the form of domain walls or, in the case of non-observation, can improve the existing bounds on non-gravitational couplings by up to eleven orders of magnitude.

%
%

\subsection{ Investigating quantum correlations and testing Bell inequalities for different gravitational
potentials and relative velocities - Measurement concept}\index{Entangle:Measurements}

For the measurement of quantum correlations and test of the violation of the Bell Inequality,  entangled photon pairs must be generated on one satellite. One photon of the pair is measured locally after a suitable delay. The other photon is sent to the other satellite using the same optical channel realized for the GW detector and at distances up to 5000-30000 km (depending on the optical losses in the link modeled according to the telescope configuration and size), where it is measured.
The crucial parameter are optical losses: typical entangled sources may produce pairs of entangled photons at a rate of $10$ to $100$ Mcounts/s. Optical losses should be kept as low as possible in order to distinguish a true coincidence from the so called accidental coincidences due to dark counts of the detectors and background radiation.

Different orbits can be useful for testing Bell's inequality and other fundamental properties of quantum physics: satellites at large distance will provide a test of the maximal distance of entanglement; satellites in relative motion will test wavefunction collapse; satellites in different gravitational potential will test Quantum Field Theory in curved spacetime. 
The suitable orbits belong to the class of Highly-Elliptical-Orbits in order to investigate the relevant effects for a large difference in gravitational potential and relative velocity.


Concerning experimental generation of entanglement,  polarization entanglement is the preferred choice due to its realization simplicity and easy measurement setup (essentially waveplates, polarizers
and single photon detectors) \cite{vall15prl}. 
Polarization entanglement requires the calibration of the relative reference frames of the transmitter and receiver satellites, in order to have a common identification of the linear polarizations.

However, other degrees of freedom such as time-bin can be also considered for entanglement generation. As recently demonstrated~\cite{vall15prl},
time-bin encoding can be transmitted over long distances in free space: in \cite{vall16prl} it was shown that time-bin encoding
survives the passage through the atmosphere, and the reflection by a moving satellite.

For such a degree of freedom, it is necessary to take into account the gravitational redshift induced by different gravitational potentials.
Indeed, the propagation between two locations
with different gravitational potentials ${\phi}$ will modify the phase relation of the entangled state.
If the receiver is located at an altitude that is different by the quantity $h$ with respect to the transmitter, the redshift of the angular frequency $\omega$ is given by:
\begin{equation}
\Delta \omega = \frac{\Delta \phi}{c^2} \omega  \approx \frac{g h}{c^2} \omega
\end{equation}
where $g$ is the free fall acceleration on the Earth's surface \cite{ride12cqg}. Such redshift is converted into a phase change in
the time-bin entangled state.

We note that this experimental scheme may directly link the quantum correlations to an effect of General Relativity.  
As described in \cite{ride12cqg,zych12cqg}, 
the above described scheme represents an optical version of the Colella, Overhauser, and Werner \cite{cole75prl}
and it can probe quantum mechanics in curved space-time. 

The measurement of time-bin entanglement requires a modification of the entanglement source and the measurement setup, since two stable unbalanced Mach-Zehnder interferometers are required (one at the source and the other at the receiver).
Moreover, in order to achieve the high visibility required for Bell-inequality violation, coupling into single mode fiber could be necessary~\cite{marcikic2002pra,vedovato2018prl}.

In conclusion, as far as testing Bell's inequalities is concerned, SAGE mission configuration will allow (1) to test quantum entanglement and thus quantum mechanics over unprecedented distances; any violation of the Bell inequality  will certify the presence of large-distance entanglement. (2) To address the fundamental problem of the unification of quantum mechanics and general relativity; by testing quantum correlations in a not-flat and varying gravitational background, SAGE  will indeed test Quantum Field Theory in curved spacetime, the first step towards a Quantum Gravity Theory. (3) To test alternative theories of quantum-decorrelation due to distance/time and gravitation potential and get insight on the wavefunction collapse.

\subsection{Defining an ultraprecise frame of reference for Earth and Space, comparing terrestrial clocks,
using clocks and links between satellites for optical VLBI in Space - Measurement concepts}\index{Reference:Measurements}
Two  SAGE satellites, or preferably three, would be operated in an optimized orbit, for example in a geostationary one. The satellites are drag-free and a laser or microwave link continuously measures the distance between the satellites. A reduced number of satellite pairs would also be a reasonable implementation.

The position of a receiver on Earth or in Space can be determined by receiving microwave or optical ranging signals from the SAGE satellites thus measuring the satellite-to-receiver distances. An additional method for obtaining kinematic information on the receiver is to measure its velocity. This is done by receiving the timing signals of the SAGE atomic clock. The Doppler shift between the SAGE clock signals and the on-board atomic clock yields the receiver velocity along the connecting line.


To perform long baseline interferometry at optical or infrared wavelength, a narrow spectral region at each telescope is coherently detected using an ultrastable optical local oscillator derived from an optical clock, which is converted to the receiving frequency using an optical frequency comb. To ensure LO coherence between the satellites, the local oscillator wave of one telescope is transferred to the second satellite to lock its local oscillator there. The distance between the two satellites must be continuously monitored by sending back the LO to the first satellite. The downconverted optical signal from both satellites is then brought together and processed to provide information on the astronomical source with ultrahigh angular resolution. Further study is required to determine whether the satellites must be of the drag-free type.






\section{Technology readiness level and roadmap for the SAGE mission}

Key components and technologies for the SAGE mission include: atomic physics package, optical bench, clock laser, optical links, drag-free and attitude control.

{Atomic physics package}:  
Compact and transportable Sr atomic physics packages have been recently developed for optical atomic clock programs as, for example, for the ESA SOC mission on ISS \cite{Poli2014,Bongs2015,Origlia2018}.  These systems contain the required laser, opto-mechanics and vacuum hardware to produce laser cooled ensembles. While they have not been space qualified, several key subsystems and components are commercially available or have already initiated the process that will bring them to flight readiness.
Operation of the atomic physics package requires several lasers. These lasers are available at the correct wavelengths and with the required characteristics from a growing number of vendors and routinely used for applications in ground-based optical atomic clocks systems. Such lasers need to be space qualified. 

{Optical bench}:   
The optical bench consists of the telescope and the opto-mechanics required for the heterodyne laser system. The telescope exit diameter is $\sim$ 30 cm. The optical path length through the telescope and beam splitter needs to be stable at a level below the specified atomic shot-noise readout level for the 2 photon recoil interferometer. This requirement sets the limit to the allowable thermal fluctuations of the telescope and beam splitter elements due to thermally induced changes in the index of refraction and expansion. It does not appear that a low thermal expansion bench is necessary, although this requires further definition and study.

{Clock laser}:  
The clock laser system operates at the Sr 698 nm clock transition. The master oscillator can be based on an external-cavity diode lasers locked to a stable cavity and followed by a Ti:Sapphire or semiconductor CW amplifier. Optical cavities for space operation are under development in several programs; environmental tests are ongoing to progress towards flight readiness. Several companies are currently working on compact and robust laser systems able to operate in harsh environments, eventually evaluating and upgrading them for space operation.

{Optical links}: 
Studies have been performed on feasibility and performances of space-space and space-Earth optical links in the frame of other space projects, notably ACES and LISA. The microwave link developed for the ACES mission is about to demonstrate flight level maturity. Several studies are on-going in the LISA mission to advance the technology for space-space optical links with performance compatible with SAGE.

{Drag-free and Attitude control}:  
The technology readiness level of drag-free and attitude control systems has now reached flight maturity. Several packages have already been flown in space and demonstrated excellent performance. The technology developed for the GOCE (Gravity field and steady-state Ocean Circulation Explorer) mission \cite{GOCE}  might already be sufficient for the SAGE instrument. MICROSCOPE has recently demonstrated control of non-gravitational accelerations at the level of $1.5\times10^{-12}$~ms$^{-2}$Hz$^{-1/2}$ \cite{Touboul2017}. A more complex system for drag-free control has  demonstrated a residual acceleration noise of $5.2\times10^{-15}$~ms$^{-2}$Hz$^{-1/2}$in the LISA Pathfinder space mission \cite{Armano2016}.
%
%


A determination of the mission costs is beyond the scope of this paper which is focused on the innovative idea of using atoms for precisely sensing the space-time structure of the Universe. Cost evaluation will be the result of a feasibility study. It will depend on the number of satellites (2 or 3), the orbit, the distance between the satellites, the characteristics of the satellite-to-satellite and satellite-to-Earth optical links. However, it can certainly be stated that the need of a multi-satellite configuration, the required orbit, and the complexity of the instrumentation set SAGE in the range corresponding to a large-scale space  mission.





While a precise definition of the roadmap to the mission is beyond the scope of this paper, a preliminary time plan can be drawn in analogy to the one for the LISA mission. 
Taking as a reference the final report of the ESA Gravitational Observatory Advisory Team (GOAT) \cite{GOAT} and considering a longer initial phase due to the lower technology readiness level of the SAGE instruments, we estimate a duration of not less than 20 years from the beginning of the study and the development of the most critical technologies to launch. A SAGE-type mission would then fit the  ESA Voyage 2050 Science plan \cite{Voyage2050}.

\section{Conclusions}{\label{Conclusions}

We presented the  idea and the main scientific and technological aspects for the  Space  Atomic  Gravity  Explorer  (SAGE)  proposed space mission. Using  quantum sensors based on ultracold strontium atoms, gravitational waves, dark matter, and other fundamental aspects of gravity would be investigated as well as the connection between gravitational physics and quantum physics. Possible applications have also been discussed.

\section*{Acknowledgments}
G.~M.~T., S.~C., N.~P., and G.~R. acknowledge financial support by  INFN and the Italian Ministry of Education, University and Research (MIUR)
under the PRIN 2015 project ''Interferometro Atomico Avanzato
per Esperimenti su Gravit\`a e Fisica Quantistica e
Applicazioni alla Geofisica''. N. P. acknowledges financial support from European Research Council,  Grant No. 772126  (TICCTOCGRAV).
D.~S. is grateful for personal funding by the Federal Ministry of Education and Research (BMBF) through the funding program Photonics Research Germany under contract number 13N14875.
A.~R.\ and W.~P.~S. acknowledge financial support by the German Space Agency (DLR) with funds provided by the Federal Ministry of Economics and Energy (BMWi) under Grants No.~50WM1556 (QUANTUS IV) and 50WM1956 (QUANTUS V).
W.v.K. acknowledges  support from the Horizon 2020 ATTRACT grant CEMIC.
F.~S. thanks the Netherlands Organisation for Scientific Research (NWO) for funding through Vici grant No. 680-47-619 and the European Union for funding through the Horizon 2020 research and innovation programme under grant agreement No. 820404 (iqClock project). 
A.D. acknowledges partial support from the U.S. National Science Foundation. 
The members of AtomQT acknowledge support by the COST Action CA16221.
M.-S.~Z. acknowledges partial support from the Strategic Priority Research Program 
of the Chinese Academy of Sciences under grant No. XDB21010100.
W.~E., N.~G., E.~R., and C.~S. acknowledge financial support by ``Nieders\"achsisches Vorab'' through the ``Quantum- and Nano- Metrology (QUANOMET)'' initiative within the project QT3.
S.~S. thanks ESA for supporting the development of the mission I-SOC through a design study.



\begin{thebibliography}{100}

\bibitem{ESA2016}
{\em https://www.cosmos.esa.int/web/new-scientific-ideas}.

\bibitem{Cronin2009}
Alexander~D Cronin, J{\"o}rg Schmiedmayer, and David~E Pritchard.
\newblock {Optics and interferometry with atoms and molecules}.
\newblock {\em Reviews of Modern Physics}, 81(3):1051--1129, July 2009.

\bibitem{Tino2014}
G.~M. Tino and M.~A. Kasevich, editors.
\newblock {\em Atom Interferometry}.
\newblock SIF and IOS Press, 2014.

\bibitem{Poli2013}
N~Poli, C~W Oates, P~Gill, and G~M Tino.
\newblock {Optical atomic clocks}.
\newblock {\em La Rivista del Nuovo Cimento}, 36:555--624, December 2013.

\bibitem{Ludlow2015}
Andrew~D Ludlow, Martin~M Boyd, Jun Ye, E~Peik, and P~O Schmidt.
\newblock {Optical atomic clocks}.
\newblock {\em Reviews of Modern Physics}, 87(2):637--701, June 2015.

\bibitem{Safronova2018}
M.~S. Safronova, D.~Budker, D.~DeMille, Derek F.~Jackson Kimball,
  A.~Derevianko, and Charles~W. Clark.
\newblock {Search for new physics with atoms and molecules}.
\newblock {\em {Rev.Mod.Phys.}}, {90}({2}), {2018}.

\bibitem{Cacciapuoti2009}
L.~Cacciapuoti and Ch. Salomon.
\newblock {Space clocks and fundamental tests: The ACES experiment}.
\newblock {\em {European Physical Journal-Special Topics}}, {172}:{57--68},
  {2009}.

\bibitem{Laurent2015}
Philippe Laurent, Didier Massonnet, Luigi Cacciapuoti, and Christophe Salomon.
\newblock {The ACES/PHARAO space mission}.
\newblock {\em {Comptes Rendus Physique}}, {16}({5}):{540--552}, {JUN} {2015}.

\bibitem{Liu2018}
Liang Liu, De-Sheng Lu, Wei-Biao Chen, Tang Li, Qiu-Zhi Qu, Bin Wang, Lin Li,
  Wei Ren, Zuo-Ren Dong, Jian-Bo Zhao, Wen-Bing Xia, Xin Zhao, Jing-Wei Ji,
  Mei-Feng Ye, Yan-Guang Sun, Yuan-Yuan Yao, Dan Song, Zhao-Gang Liang,
  Shan-Jiang Hu, Dun-He Yu, Xia Hou, Wei Shi, Hua-Guo Zang, Jing-Feng Xiang,
  Xiang-Kai Peng, and Yu-Zhu Wang.
\newblock {In-orbit operation of an atomic clock based on laser-cooled
  $^{87}$Rb atoms}.
\newblock {\em Nature Com.}, 9:2760, July 24 2018.

\bibitem{Bongs2015}
K.~Bongs, Y.~Singh, L.~Smith, W.~He, O.~Kock, D.~Swierad, J.~Hughes,
  S.~Schiller, S.~Alighanbari, S.~Origlia, S.~Vogt, U.~Sterr, Ch. Lisdat,
  R.~Le~Targat, J.~Lodewyck, D.~Holleville, B.~Venon, S.~Bize, G.~P. Barwood,
  P.~Gill, I.~R. Hill, Y.~B. Ovchinnikov, N.~Poli, G.~M. Tino, J.~Stuhler,
  W.~Kaenders, and {the SOC2 team}.
\newblock Development of a strontium optical lattice clock for the {SOC}
  mission on the {ISS}.
\newblock {\em Comptes Rendus Physique}, {16}({5}):{553--564}, {2015}.

\bibitem{ISOC-ESR}
S.~Schiller and L.~Cacciapuoti.
\newblock {I-SOC} {Scientific Requirements}.
\newblock SCI-ESA-HRE-ESR-ISOC, 2017.

\bibitem{Origlia2018}
S.~Origlia, M.~S. Pramod, S.~Schiller, Y.~Singh, K.~Bongs, R.~Schwarz,
  A.~Al-Masoudi, S.~D\"orscher, S.~Herbers, S.~H\"afner, U.~Sterr, and Ch.
  Lisdat.
\newblock Towards an optical clock for space: Compact, high-performance optical
  lattice clock based on bosonic atoms.
\newblock {\em Phys. Rev. A}, 98:053443, Nov 2018.

\bibitem{Tino2007a}
G.~M. Tino, L.~Cacciapuoti, K.~Bongs, Ch.~J. Borde, P.~Bouyer, H.~Dittus,
  W.~Ertmer, A.~Goerlitz, M.~Inguscio, A.~Landragin, P.~Lemonde, C.~Lammerzahl,
  A.~Peters, E.~Rasel, J.~Reichel, C.~Salomon, S.~Schiller, W.~Schleich,
  K.~Sengstock, U.~Sterr, and M.~Wilkens.
\newblock {Atom interferometers and optical atomic clocks: New quantum sensors
  for fundamental physics experiments in space}.
\newblock {\em {Nuclear Physics B -Proceedings Supplements}}, {166}:{159--165},
  {2007}.

\bibitem{Sorrentino2010}
Fiodor Sorrentino, Kai Bongs, Philippe Bouyer, Luigi Cacciapuoti, Marella
  de~Angelis, Hansjoerg Dittus, Wolfgang Ertmer, A.~Giorgini, J.~Hartwig,
  Matthias Hauth, Sven Herrmann, Massimo Inguscio, Endre Kajari, Thorben~T.
  Koenemann, Claus Laemmerzahl, Arnaud Landragin, Giovanni Modugno,
  Frank~Pereira dos Santos, Achmin Peters, Marco Prevedelli, Ernst~M. Rasel,
  Wolfgang~P. Schleich, Malte Schmidt, Alexander Senger, Klaus Sengstock,
  Guillaume Stern, Guglielmo~Maria Tino, and Reinhold Walser.
\newblock {A Compact Atom Interferometer for Future Space Missions}.
\newblock {\em {Microgravity Science and Technology}}, {22}({4}):{551--561},
  {2010}.

\bibitem{Wolf2009}
P.~Wolf, Ch.~J. Borde, A.~Clairon, L.~Duchayne, A.~Landragin, P.~Lemonde,
  G.~Santarelli, W.~Ertmer, E.~Rasel, F.~S. Cataliotti, M.~Inguscio, G.~M.
  Tino, P.~Gill, H.~Klein, S.~Reynaud, C.~Salomon, E.~Peik, O.~Bertolami,
  P.~Gil, J.~Paramos, C.~Jentsch, U.~Johann, A.~Rathke, P.~Bouyer,
  L.~Cacciapuoti, D.~Izzo, P.~De~Natale, B.~Christophe, P.~Touboul, S.~G.
  Turyshev, J.~Anderson, M.~E. Tobar, F.~Schmidt-Kaler, J.~Vigue, A.~A. Madej,
  L.~Marmet, M.~C. Angonin, P.~Delva, P.~Tourrenc, G.~Metris, H.~Mueller,
  R.~Walsworth, Z.~H. Lu, L.~J. Wang, K.~Bongs, A.~Toncelli, M.~Tonelli,
  H.~Dittus, C.~Laemmerzahl, G.~Galzerano, P.~Laporta, J.~Laskar, A.~Fienga,
  F.~Roques, and K.~Sengstock.
\newblock {Quantum physics exploring gravity in the outer solar system: the
  SAGAS project}.
\newblock {\em {Experimental Astronomy}}, {23}({2}):{651--687}, {2009}.

\bibitem{Aguilera2014}
D.~Aguilera, H.~Ahlers, B.~Battelier, A.~Bawamia, A.~Bertoldi, R.~Bondarescu,
  K.~Bongs, P.~Bouyer, C.~Braxmaier, L.~Cacciapuoti, C.~Chaloner, M.~Chwalla,
  W.~Ertmer, M.~Franz, N.~Gaaloul, M.~Gehler, D.~Gerardi, L.~Gesa,
  N.~Guerlebeck, J.~Hartwig, M.~Hauth, O.~Hellmig, W.~Herr, S.~Herrmann,
  A.~Heske, A.~Hinton, P.~Ireland, P.~Jetzer, U.~Johann, M.~Krutzik,
  A.~Kubelka, C.~Laemmerzahl, A.~Landragin, I.~Lloro, D.~Massonnet, I.~Mateos,
  A.~Milke, M.~Nofrarias, M.~Oswald, A.~Peters, K.~Posso-Trujillo, E.~Rasel,
  E.~Rocco, A.~Roura, J.~Rudolph, W.~Schleich, C.~Schubert, T.~Schuldt,
  S.~Seidel, K.~Sengstock, C.~F. Sopuerta, F.~Sorrentino, D.~Summers, G.~M.
  Tino, C.~Trenkel, N.~Uzunoglu, W.~von Klitzing, R.~Walser, T.~Wendrich,
  A.~Wenzlawski, P.~Wessels, A.~Wicht, E.~Wille, M.~Williams, P.~Windpassinger,
  and N.~Zahzam.
\newblock {STE-QUEST-test of the universality of free fall using cold atom
  interferometry}.
\newblock {\em Cassical and Quantum Gravity}, {31}({15}), 2014.

\bibitem{Altschul2015}
Brett Altschul, Quentin~G. Bailey, Luc Blanchet, Kai Bongs, Philippe Bouyer,
  Luigi Cacciapuoti, Salvatore Capozziello, Naceur Gaaloul, Domenico Giulini,
  Jonas Hartwig, Luciano Iess, Philippe Jetzer, Arnaud Landragin, Ernst Rasel,
  Serge Reynaud, Stephan Schiller, Christian Schubert, Fiodor Sorrentino, Uwe
  Sterr, Jay~D. Tasson, Guglielmo~M. Tino, Philip Tuckey, and Peter Wolf.
\newblock Quantum tests of the {Einstein} {Equivalence} {Principle} with the
  {STE-QUEST} space mission.
\newblock {\em Advances in Space Research}, 55(1):501 -- 524, 2015.

\bibitem{Canuel2018}
B.~Canuel, A.~Bertoldi, L.~Amand, E.~Pozzo di~Borgo, T.~Chantrait,
  C.~Danquigny, M.~Dovale Alvarez, B.~Fang, A.~Freise, R.~Geiger, J.~Gillot,
  S.~Henry, J.~Hinderer, D.~Holleville, J.~Junca, G.~Lefevre, M.~Merzougui,
  N.~Mielec, T.~Monfret, S.~Pelisson, M.~Prevedelli, S.~Reynaud, I.~Riou,
  Y.~Rogister, S.~Rosat, E.~Cormier, A.~Landragin, W.~Chaibi, S.~Gaffet, and
  P.~Bouyer.
\newblock {Exploring gravity with the MIGA large scale atom interferometer}.
\newblock {\em {Scientific Reports}}, {8}, {2018}.

\bibitem{MAGIS100}
{\em https://qis.fnal.gov/magis-100/}.

\bibitem{Zhan2019}
Ming-Sheng Zhan, Jin Wang, Wei-Tou Ni, Dong-Feng Gao, Gang Wang, Ling-Xiang He,
  Run-Bing Li, Lin Zhou, Xi~Chen, Jia-Qi Zhong, Biao Tang, Zhan-Wei Yao, Lei
  Zhu, Zong-Yuan Xiong, Si-Bin Lu, Geng-Hua Yu, Qun-Feng Cheng, Min Liu,
  Yu-Rong Liang, Peng Xu, Xiao-Dong He, Min Ke, Zheng Tan, and Jun Luo.
\newblock {ZAIGA: Zhaoshan Long-baseline Atom Interferometer Gravitation
  Antenna}.
\newblock {\em Int. J. Mod. Phys. D}, 28:1940005, 2019.

\bibitem{Becker2018}
Dennis Becker, Maike~D. Lachmann, Stephan~T. Seidel, Holger Ahlers, Aline~N.
  Dinkelaker, Jens Grosse, Ortwin Hellmig, Hauke Muentinga, Vladimir Schkolnik,
  Thijs Wendrich, Andre Wenzlawski, Benjamin Weps, Robin Corgier, Tobias Franz,
  Naceur Gaaloul, Waldemar Herr, Daniel Luedtke, Manuel Popp, Sirine Amri,
  Hannes Duncker, Maik Erbe, Anja Kohfeldt, Andre Kubelka-Lange, Claus
  Braxmaier, Eric Charron, Wolfgang Ertmer, Markus Krutzik, Claus Laemmerzahl,
  Achim Peters, Wolfgang~P. Schleich, Klaus Sengstock, Reinhold Walser, Andreas
  Wicht, Patrick Windpassinger, and Ernst~M. Rasel.
\newblock {Space-borne Bose-Einstein condensation for precision
  interferometry}.
\newblock {\em {Nature}}, {562}({7727}):{391}, {2018}.

\bibitem{GOAT}
{ESA} {G}ravitational {O}bservatory {A}dvisory {T}eam -
  www.cosmos.esa.int/web/goat.

\bibitem{Gao2018}
Dongfeng Gao, Jin Wang, and Mingsheng Zhan.
\newblock {Atomic Interferometric Gravitational-wave Space Observatory
  (AIGSO)}.
\newblock {\em Commun. Theor. Phys.}, 69:37--42, 2018.

\bibitem{Wang2019}
Gang Wang, Dongfeng Gao, Wei-Tou Ni, Jin Wang, and Mingsheng Zhan.
\newblock {Orbit Design for Space Atom-Interferometer AIGSO}.
\newblock {\em Int. J. Mod. Phys. D}, 28:1940004, 2019.

\bibitem{abbott2016observation}
B.P.~Abbott et~al. LIGO and Virgo Collaborations.
\newblock {Observation of gravitational waves from a binary black hole merger}.
\newblock {\em Phys. Rev. Lett.}, 116(6):061102, 2016.

\bibitem{abbott2017b}
B.P.~Abbott et~al. LIGO and Virgo Collaborations.
\newblock Gw170814: A three-detector observation of gravitational waves from a
  binary black hole coalescence.
\newblock {\em Phys. Rev. Lett.}, 119:141101, 2017.

\bibitem{Armano2016}
M.~Armano et~al.
\newblock {Sub-Femto-g Free Fall for Space-Based Gravitational Wave
  Observatories: {LISA} {P}athfinder Results}.
\newblock {\em Phys. Rev. Lett.}, 116:231101, 2016.

\bibitem{L3_final_report}
{The {ESA-L3} Gravitational Wave Mission - Final Report}.
\newblock Technical report,
  http://sci.esa.int/cosmic-vision/57910-goat-final-report-on-the-esa-l3-gravitational-wave-mission/,
  2016.

\bibitem{Tino2007b}
G.~M. Tino and F.~Vetrano.
\newblock {Is it possible to detect gravitational waves with atom
  interferometers?}
\newblock {\em Classical and Quantum Gravity}, 24(9):2167--2178, 2007.

\bibitem{Tino2011}
Special issue on ``{G}ravitational waves detection with atom interferometry''.
\newblock In G.~M. Tino, F.~Vetrano, and C.~L{\"a}mmerzahl, editors, {\em Gen.
  Relativ. Gravit.}, volume~43, page 1901. 2011.

\bibitem{Dimopoulos2008}
S.~Dimopoulos, P.~Graham, J.~Hogan, M.~Kasevich, and S.~Rajendran.
\newblock {Atomic gravitational wave interferometric sensor}.
\newblock {\em Phys. Rev. D}, 78(12):122002, 2008.

\bibitem{Kolkowitz2016}
S.~Kolkowitz, I.~Pikovski, N.~Langellier, M.~D. Lukin, R.~L. Walsworth, and
  J.~Ye.
\newblock Gravitational wave detection with optical lattice atomic clocks.
\newblock {\em Phys. Rev. D}, 94:124043, 2016.

\bibitem{Ferrari2006b}
G.~Ferrari, N.~Poli, F.~Sorrentino, and G.~M. Tino.
\newblock {Long-Lived Bloch Oscillations with Bosonic Sr Atoms and Application
  to Gravity Measurement at the Micrometer Scale}.
\newblock {\em Phys.\ Rev.\ Lett.}, 97:060402, 2006.

\bibitem{Mazzoni2015}
T.~Mazzoni, X.~Zhang, R.~Del Aguila, L.~Salvi, N.~Poli, and G.~M. Tino.
\newblock {Large-momentum-transfer {B}ragg interferometer with strontium
  atoms}.
\newblock {\em Physical Review A}, 92(5):053619, 2015.

\bibitem{Hu2017}
Liang Hu, Nicola Poli, Leonardo Salvi, and Guglielmo~M Tino.
\newblock {Atom Interferometry with the Sr Optical Clock Transition}.
\newblock {\em Physical Review Letters}, 119(26):555--5, 2017.

\bibitem{Loriani2019}
S~Loriani, D~Schlippert, C~Schubert, S~Abend, H~Ahlers, W~Ertmer, J~Rudolph,
  J~M Hogan, M~A Kasevich, E~M Rasel, and N~Gaaloul.
\newblock Atomic source selection in space-borne gravitational wave detection.
\newblock {\em New Journal of Physics}, 21(6):063030, jun 2019.

\bibitem{Norcia2017}
Matthew~A. Norcia, Julia R.~K. Cline, and James~K. Thompson.
\newblock Role of atoms in atomic gravitational-wave detectors.
\newblock {\em Phys. Rev. A}, 96:042118, Oct 2017.

\bibitem{Lada2006}
C.~J. Lada.
\newblock {Stellar Multiplicity and the Initial Mass Function: Most Stars Are
  Single}.
\newblock {\em ApJ}, 640(1):L63, 2006.

\bibitem{Abbott2017observation}
B.P.~Abbott et~al. LIGO and Virgo Collaborations.
\newblock Gw170817: Observation of gravitational waves from a binary neutron
  star inspiral.
\newblock {\em Phys. Rev. Lett.}, 119:161101, 2017.

\bibitem{shah2014}
Sweta Shah and Gijs Nelemans.
\newblock {Constraining Parameters of White-Dwarf Binaries Using
  Gravitational-Wave and Electromagnetic Observations}.
\newblock {\em {Astrophysical Journal}}, {790}({2}), {Aug 1} {2014}.

\bibitem{AdV}
F.~Acernese et~al Virgo~Collaboration.
\newblock { Advanced Virgo: a second generation interferometric gravitational
  wave detector}.
\newblock {\em Classical and Quantum Gravity}, 32:024001, 2015.

\bibitem{AdL}
{\em https://dcc.ligo.org/LIGO-T1800044/public}.

\bibitem{Feng2010}
Jonathan~L Feng.
\newblock {Dark Matter Candidates from Particle Physics and Methods of
  Detection}.
\newblock {\em Ann. Rev. Astro. Astrophys.}, 48(1):495--545, aug 2010.

\bibitem{Akerib:2013tjd}
D.~S. Akerib, H.~M. Ara{\'{u}}jo, X.~Bai, A.~J. Bailey, J.~Balajthy,
  S.~Bedikian, E.~Bernard, A.~Bernstein, A.~Bolozdynya, A.~Bradley, D.~Byram,
  S.~B. Cahn, M.~C. Carmona-Benitez, C.~Chan, J.~J. Chapman, A.~A. Chiller,
  C.~Chiller, K.~Clark, T.~Coffey, A.~Currie, A.~Curioni, S.~Dazeley, L.~{De
  Viveiros}, A.~Dobi, J.~Dobson, E.~M. Dragowsky, E.~Druszkiewicz, B.~Edwards,
  C.~H. Faham, S.~Fiorucci, C.~Flores, R.~J. Gaitskell, V.~M. Gehman, C.~Ghag,
  K.~R. Gibson, M.~G~D Gilchriese, C.~Hall, M.~Hanhardt, S.~A. Hertel, M.~Horn,
  D.~Q. Huang, M.~Ihm, R.~G. Jacobsen, L.~Kastens, K.~Kazkaz, R.~Knoche,
  S.~Kyre, R.~Lander, N.~A. Larsen, C.~Lee, D.~S. Leonard, K.~T. Lesko,
  A.~Lindote, M.~I. Lopes, A.~Lyashenko, D.~C. Malling, R.~Mannino, D.~N.
  McKinsey, D.~M. Mei, J.~Mock, M.~Moongweluwan, J.~Morad, M.~Morii, A.~St~J
  Murphy, C.~Nehrkorn, H.~Nelson, F.~Neves, J.~A. Nikkel, R.~A. Ott,
  M.~Pangilinan, P.~D. Parker, E.~K. Pease, K.~Pech, P.~Phelps, L.~Reichhart,
  T.~Shutt, C.~Silva, W.~Skulski, C.~J. Sofka, V.~N. Solovov, P.~Sorensen,
  T.~Stiegler, K.~O'Sullivan, T.~J. Sumner, R.~Svoboda, M.~Sweany, M.~Szydagis,
  D.~Taylor, B.~Tennyson, D.~R. Tiedt, M.~Tripathi, S.~Uvarov, J.~R. Verbus,
  N.~Walsh, R.~Webb, J.~T. White, D.~White, M.~S. Witherell, M.~Wlasenko,
  F.~L~H Wolfs, M.~Woods, and C.~Zhang.
\newblock {First results from the LUX dark matter experiment at the Sanford
  underground research facility}.
\newblock {\em Phys. Rev. Lett.}, 112, 2014.

\bibitem{Marsh2016}
David~J.E. Marsh.
\newblock {Axion cosmology}.
\newblock {\em Physics Reports}, 643:1--79, 2016.

\bibitem{Hu2000}
Wayne Hu, Rennan Barkana, and Andrei Gruzinov.
\newblock {Fuzzy Cold Dark Matter: The Wave Properties of Ultralight
  Particles}.
\newblock {\em Phys. Rev. Lett.}, 85(6):1158--1161, aug 2000.

\bibitem{Marsh2014}
David J~E Marsh and Joseph Silk.
\newblock {A model for halo formation with axion mixed dark matter}.
\newblock {\em Mon. Not. R. Astron. Soc.}, 2014.

\bibitem{Hui2017}
Lam Hui, Jeremiah~P Ostriker, Scott Tremaine, and Edward Witten.
\newblock {Ultralight scalars as cosmological dark matter}.
\newblock {\em Phys. Rev. D}, 2017.

\bibitem{Arvanitaki2010}
Asimina Arvanitaki, Savas Dimopoulos, Sergei Dubovsky, Nemanja Kaloper, and
  John March-Russell.
\newblock {String axiverse}.
\newblock {\em Phys. Rev. D}, 81(12):123530, jun 2010.

\bibitem{Tilburg2015}
Ken Van~Tilburg, Nathan Leefer, Lykourgos Bougas, and Dmitry Budker.
\newblock Search for ultralight scalar dark matter with atomic spectroscopy.
\newblock {\em Phys. Rev. Lett.}, 115:011802, Jun 2015.

\bibitem{Hees2016}
A.~Hees, J.~Gu\'ena, M.~Abgrall, S.~Bize, and P.~Wolf.
\newblock Searching for an oscillating massive scalar field as a dark matter
  candidate using atomic hyperfine frequency comparisons.
\newblock {\em Phys. Rev. Lett.}, 117:061301, Aug 2016.

\bibitem{Hees2018}
Aur\'elien Hees, Olivier Minazzoli, Etienne Savalle, Yevgeny~V. Stadnik, and
  Peter Wolf.
\newblock Violation of the equivalence principle from light scalar dark matter.
\newblock {\em Phys. Rev. D}, 98:064051, Sep 2018.

\bibitem{Wcislo2018}
P.~Wcis{\l}o, P.~Ablewski, K.~Beloy, S.~Bilicki, M.~Bober, R.~Brown, R.~Fasano,
  R.~Ciury{\l}o, H.~Hachisu, T.~Ido, J.~Lodewyck, A.~Ludlow, W.~McGrew,
  P.~Morzy{\'n}ski, D.~Nicolodi, M.~Schioppo, M.~Sekido, R.~Le~Targat, P.~Wolf,
  X.~Zhang, B.~Zjawin, and M.~Zawada.
\newblock New bounds on dark matter coupling from a global network of optical
  atomic clocks.
\newblock {\em Science Advances}, 4(12), 2018.

\bibitem{Roberts2017-GPS-DM}
Benjamin~M. Roberts, Geoffrey Blewitt, Conner Dailey, Mac Murphy, Maxim
  Pospelov, Alex Rollings, Jeff Sherman, Wyatt Williams, and Andrei Derevianko.
\newblock {Search for domain wall dark matter with atomic clocks on board
  global positioning system satellites}.
\newblock {\em Nature Comm.}, 8(1):1195, dec 2017.

\bibitem{Derevianko2014}
A.~Derevianko and M.~Pospelov.
\newblock {Hunting for topological dark matter with atomic clocks}.
\newblock {\em Nature Phys.}, 10:933, 2014.

\bibitem{Derevianko2016a}
Andrei Derevianko.
\newblock {Detecting dark-matter waves with a network of precision-measurement
  tools}.
\newblock {\em Phys. Rev. A}, 97(4):042506, apr 2018.

\bibitem{Arvanitaki2015}
Asimina Arvanitaki, Savas Dimopoulos, and Ken {Van Tilburg}.
\newblock {Sound of Dark Matter: Searching for Light Scalars with Resonant-Mass
  Detectors}.
\newblock {\em Phys. Rev. Lett.}, 116(3):031102, jan 2016.

\bibitem{Wcislo-clock-network-2018}
P.~Wcis{\l}o, P.~Ablewski, K.~Beloy, S.~Bilicki, M.~Bober, R.~Brown, R.~Fasano,
  R.~Ciury{\l}o, H.~Hachisu, T.~Ido, J.~Lodewyck, A.~Ludlow, W.~McGrew,
  P.~Morzy{\'n}ski, D.~Nicolodi, M.~Schioppo, M.~Sekido, R.~Le~Targat, P.~Wolf,
  X.~Zhang, B.~Zjawin, and M.~Zawada.
\newblock New bounds on dark matter coupling from a global network of optical
  atomic clocks.
\newblock {\em Science Advances}, 4(12):eaau4869, 2018.

\bibitem{Wcislo2016}
P.~Wcis{\l}o, P.~Morzy\'nski, M.~Bober, A.~Cygan, D.~Lisak, R.~Ciury{\l}o, and
  M.~Zawada.
\newblock Searching for topological defect dark matter with optical atomic
  clocks.
\newblock {\em Nat. Astron.}, 1:0009, 2016.

\bibitem{Derevianko2016}
Andrei Derevianko.
\newblock Atomic clocks and dark-matter signatures.
\newblock {\em Journal of Physics: Conference Series}, 723:012043, 2016.

\bibitem{Touboul2017}
Pierre Touboul, Gilles M\'etris, Manuel Rodrigues, Yves Andr\'e, Quentin Baghi,
  Jo\"el Berg\'e, Damien Boulanger, Stefanie Bremer, Patrice Carle, Ratana
  Chhun, Bruno Christophe, Valerio Cipolla, Thibault Damour, Pascale Danto,
  Hansjoerg Dittus, Pierre Fayet, Bernard Foulon, Claude Gageant, Pierre-Yves
  Guidotti, Daniel Hagedorn, Emilie Hardy, Phuong-Anh Huynh, Henri Inchauspe,
  Patrick Kayser, St\'ephanie Lala, Claus L\"ammerzahl, Vincent Lebat, Pierre
  Leseur, Fran\ifmmode \mbox{\c{c}}\else~\c{c}\fi{}oise Liorzou, Meike List,
  Frank L\"offler, Isabelle Panet, Benjamin Pouilloux, Pascal Prieur, Alexandre
  Rebray, Serge Reynaud, Benny Rievers, Alain Robert, Hanns Selig, Laura
  Serron, Timothy Sumner, Nicolas Tanguy, and Pieter Visser.
\newblock Microscope mission: First results of a space test of the equivalence
  principle.
\newblock {\em Phys. Rev. Lett.}, 119:231101, Dec 2017.

\bibitem{TOUBOUL2001}
P.~Touboul, M.~Rodrigues, G.~M{\'e}tris, and B.~Tatry.
\newblock {MICROSCOPE, testing the equivalence principle in space}.
\newblock {\em Comptes Rendus de l'Acad{\'e}mie des Sciences - Series IV -
  Physics}, 2(9):1271 -- 1286, 2001.

\bibitem{QTEST16}
J.~Williams, S.~w.~Chiow, N.~Yu, and H.~M{\"u}ller.
\newblock {Quantum test of the equivalence principle and space-time aboard the
  International Space Station}.
\newblock {\em New Journal of Physics}, 18(2):025018, 2016.

\bibitem{Will2006}
C.~M. Will.
\newblock {The Confrontation between General Relativity and Experiment}.
\newblock {\em Living Reviews in Relativity}, 9(3), 2006.

\bibitem{Wolf2016}
P.~Wolf and L.~Blanchet.
\newblock {Analysis of Sun/Moon gravitational redshift tests with the STE-QUEST
  space mission}.
\newblock {\em Classical and Quantum Gravity}, 33(3):035012, 2016.

\bibitem{Delva2018}
P.~Delva, N.~Puchades, E.~Sch\"onemann, F.~Dilssner, C.~Courde, S.~Bertone,
  F.~Gonzalez, A.~Hees, Ch. Le~Poncin-Lafitte, F.~Meynadier,
  R.~Prieto-Cerdeira, B.~Sohet, J.~Ventura-Traveset, and P.~Wolf.
\newblock Gravitational redshift test using eccentric galileo satellites.
\newblock {\em Phys. Rev. Lett.}, 121:231101, Dec 2018.

\bibitem{Herrmann2018}
Sven Herrmann, Felix Finke, Martin L\"ulf, Olga Kichakova, Dirk Puetzfeld,
  Daniela Knickmann, Meike List, Benny Rievers, Gabriele Giorgi, Christoph
  G\"unther, Hansj\"org Dittus, Roberto Prieto-Cerdeira, Florian Dilssner,
  Francisco Gonzalez, Erik Sch\"onemann, Javier Ventura-Traveset, and Claus
  L\"ammerzahl.
\newblock Test of the gravitational redshift with galileo satellites in an
  eccentric orbit.
\newblock {\em Phys. Rev. Lett.}, 121:231102, Dec 2018.

\bibitem{Schlamminger2008}
S.~Schlamminger, K.-Y. Choi, T.~A. Wagner, J.H. Gundlach, and E.~G. Adelberger.
\newblock {Test of the Equivalence Principle Using a Rotating Torsion Balance}.
\newblock {\em Phys. Rev. Lett}, 100:041101, 2008.

\bibitem{Hofmann2018}
F~Hofmann and J~Muller.
\newblock Relativistic tests with lunar laser ranging.
\newblock {\em Classical and Quantum Gravity}, 35(3):035015, 2018.

\bibitem{Doser2010}
Michael Doser and the Aegis~collaboration.
\newblock Aegis: An experiment to measure the gravitational interaction between
  matter and antimatter.
\newblock {\em Journal of Physics: Conference Series}, 199(1):012009, 2010.

\bibitem{Perez2012}
P~Perez and Y~Sacquin.
\newblock The {GBAR} experiment: gravitational behaviour of antihydrogen at
  rest.
\newblock {\em Classical and Quantum Gravity}, 29(18):184008, 2012.

\bibitem{Peters1999}
A.~{Peters}, K.~Y. {Chung}, and S.~{Chu}.
\newblock {Measurement of gravitational acceleration by dropping atoms}.
\newblock {\em Nature}, 400:849--852, Aug 1999.

\bibitem{Merlet2010}
S~Merlet, Q~Bodart, N~Malossi, A~Landragin, F~Pereira~Dos Santos, O~Gitlein,
  and L~Timmen.
\newblock Comparison between two mobile absolute gravimeters: optical versus
  atomic interferometers.
\newblock {\em Metrologia}, 47(4):L9, 2010.

\bibitem{Poli2011}
N~Poli, F~Wang, M~Tarallo, and A~Alberti.
\newblock {Precision Measurement of Gravity with Cold Atoms in an Optical
  Lattice and Comparison with a Classical Gravimeter}.
\newblock {\em PHYSICAL REVIEW LETTERS}, 106(3):038501, 2011.

\bibitem{Zhou2015}
L.~Zhou, S.~Long, B.~Tang, X.~Chen, F.~Gao, W.~Peng, W.~Duan, J.~Zhong,
  Z.~Xiong, J.~Wang, Y.~Zhang, and M.~Zhan.
\newblock {Test of Equivalence Principle at $10^{-8}$ Level by a Dual-species
  Double-diffraction Raman Atom Interferometer}.
\newblock {\em Phys. Rev. Lett}, 115:013004, 2015.

\bibitem{Tarallo2014a}
M.~G. Tarallo, T.~Mazzoni, N.~Poli, D.~V. Sutyrin, X.~Zhang, and G.~M. Tino.
\newblock {Test of Einstein Equivalence Principle for 0-Spin and
  Half-Integer-Spin Atoms: Search for Spin-Gravity Coupling Effects}.
\newblock {\em Phys. Rev. Lett.}, 113:023005, Jul 2014.

\bibitem{Schlippert2014}
D.~Schlippert, J.~Hartwig, H.~Albers, L.~L. Richardson, C.~Schubert, A.~Roura,
  W.~P. Schleich, W.~Ertmer, and E.~M. Rasel.
\newblock Quantum test of the universality of free fall.
\newblock {\em Phys. Rev. Lett.}, 112:203002, May 2014.

\bibitem{Bonnin2013}
A.~Bonnin, N.~Zahzam, Y.~Bidel, and A.~Bresson.
\newblock Simultaneous dual-species matter-wave accelerometer.
\newblock {\em Phys. Rev. A}, 88(4):043615--, 2013.

\bibitem{Fray2004}
Sebastian Fray, Cristina~Alvarez Diez, Theodor~W. H\"ansch, and Martin Weitz.
\newblock Atomic interferometer with amplitude gratings of light and its
  applications to atom based tests of the equivalence principle.
\newblock {\em Phys. Rev. Lett.}, 93(24):240404--, 2004.

\bibitem{Rosi2017}
G.~Rosi, G.~D?Amico, L.~Cacciapuoti, F.~Sorrentino, M.~Prevedelli, M.~Zych,
  ?. Brukner, and G.~M. Tino.
\newblock Quantum test of the equivalence principle for atoms in coherent
  superposition of internal energy states.
\newblock {\em Nature Comm}, 8:15529, 2017.

\bibitem{Duan2016}
Xiao-Chun Duan, Xiao-Bing Deng, Min-Kang Zhou, Ke~Zhang, Wen-Jie Xu, Feng
  Xiong, Yao-Yao Xu, Cheng-Gang Shao, Jun Luo, and Zhong-Kun Hu.
\newblock Test of the universality of free fall with atoms in different spin
  orientations.
\newblock {\em Phys. Rev. Lett.}, 117:023001, Jul 2016.

\bibitem{Roura2018}
A.~Roura.
\newblock {Gravitational redshift in quantum-clock interferometry}.
\newblock {\em arXiv:1810.06744}.

\bibitem{Rideout2012}
D.~Rideout, T.~Jennewein, G.~Amelino-Camelia, T.~F. Demarie, B.~L. Higgins,
  A.~Kempf, A.~Kent, R.~Laflamme, X.~Ma, R.~B. Mann,
  E.~Mart{\'\i}n-Mart{\'\i}nez, N.~C. Menicucci, J.~Moffat, C.~Simon,
  R.~Sorkin, L.~Smolin, and D.~R. Terno.
\newblock {Fundamental quantum optics experiments conceivable with
  satellites---reaching relativistic distances and velocities}.
\newblock {\em Classical and Quantum Gravity}, 29:224011, 2012.

\bibitem{schr35pro}
E.~Schr{\"o}dinger.
\newblock {Discussion of probability relations between separated systems}.
\newblock {\em Proceedings of the Cambridge Philosophical Society}, 31(4),
  1936.

\bibitem{bell64phy}
J.~S. Bell.
\newblock {On the Einstein-Podolsky-Rosen paradox}.
\newblock {\em Physics}, 1:195, 1964.

\bibitem{clau69prl}
J.~F. Clauser, M.~A. Horne, A.~Shimony, and R.~A. Holt.
\newblock {Proposed Experiment to Test Local Hidden-Variable Theories}.
\newblock {\em Phys. Rev. Lett.}, 23:880, 1969.

\bibitem{clau74prd}
J.~F. Clauser and M.~A. Horne.
\newblock {Experimental consequences of objective local theories}.
\newblock {\em Phys. Rev}, 10, 1974.

\bibitem{hens15nat}
B.~Hensen, H.~Bernien, A.~E. Dr{\'{e}}au, A.~Reiserer, N.~Kalb, M.~S. Blok,
  J.~Ruitenberg, R.~F.~L. Vermeulen, R.~N. Schouten, C.~Abell{\'{a}}n,
  W.~Amaya, V.~Pruneri, M.~W. Mitchell, M.~Markham, D.~J. Twitchen, D.~Elkouss,
  S.~Wehner, T.~H. Taminiau, and R.~Hanson.
\newblock {Loophole-free Bell inequality violation using electron spins
  separated by 1.3 kilometres}.
\newblock {\em Nature}, 526:682, 2015.

\bibitem{gius15prl}
Marissa Giustina, Marijn A.~M. Versteegh, S\"oren Wengerowsky, Johannes
  Handsteiner, Armin Hochrainer, Kevin Phelan, Fabian Steinlechner, Johannes
  Kofler, Jan-\AA{}ke Larsson, Carlos Abell\'an, Waldimar Amaya, Valerio
  Pruneri, Morgan~W. Mitchell, J\"orn Beyer, Thomas Gerrits, Adriana~E. Lita,
  Lynden~K. Shalm, Sae~Woo Nam, Thomas Scheidl, Rupert Ursin, Bernhard
  Wittmann, and Anton Zeilinger.
\newblock Significant-loophole-free test of {Bell's} theorem with entangled
  photons.
\newblock {\em Phys. Rev. Lett.}, 115:250401, Dec 2015.

\bibitem{shal15prl}
L.~K. Shalm, E.~Meyer-Scott, B.~G. Christensen, P.~Bierhorst, M.~A. Wayne,
  M.~J. Stevens, T.~Gerrits, S.~Glancy, D.~R. Hamel, M.~S. Allman, K.~J.
  Coakley, S.~D. Dyer, C.~Hodge, A.~E. Lita, V.~B. Verma, C.~Lambrocco,
  E.~Tortorici, A.~L. Migdall, Y.~Zhang, D.~R. Kumor, W.~H. Farr, F.~Marsili,
  M.~D. Shaw, J.~A. Stern, C.~Abell{\'{a}}n, W.~Amaya, V.~Pruneri,
  T.~Jennewein, M.~W. Mitchell, P.~G. Kwiat, J.~C. Bienfang, R.~P. Mirin,
  E.~Knill, and S.~W. Nam.
\newblock {Strong Loophole-Free Test of Local Realism}.
\newblock {\em Physical Review Letters}, 115:250402, 2015.

\bibitem{sche10pnas}
T.~Scheidl, R.~Ursin, J.~Kofler, S.~Ramelow, X.-S. Ma, T.~Herbst,
  L.~Ratschbacher, A.~Fedrizzi, N.~K. Langford, T.~Jennewein, and A.~Zeilinger.
\newblock {Violation of local realism with freedom of choice.}
\newblock {\em PNAS}, 107:19708, 2010.

\bibitem{yin2017science}
Juan Yin, Yuan Cao, Yu-Huai Li, Sheng-Kai Liao, Liang Zhang, Ji-Gang Ren,
  Wen-Qi Cai, Wei-Yue Liu, Bo~Li, Hui Dai, Guang-Bing Li, Qi-Ming Lu, Yun-Hong
  Gong, Yu~Xu, Shuang-Lin Li, Feng-Zhi Li, Ya-Yun Yin, Zi-Qing Jiang, Ming Li,
  Jian-Jun Jia, Ge~Ren, Dong He, Yi-Lin Zhou, Xiao-Xiang Zhang, Na~Wang, Xiang
  Chang, Zhen-Cai Zhu, Nai-Le Liu, Yu-Ao Chen, Chao-Yang Lu, Rong Shu,
  Cheng-Zhi Peng, Jian-Yu Wang, and Jian-Wei Pan.
\newblock Satellite-based entanglement distribution over 1200 kilometers.
\newblock {\em Science}, 356(6343):1140--1144, 2017.

\bibitem{liao2017nature}
Sheng-Kai Liao, Wen-Qi Cai, Wei-Yue Liu, Liang Zhang, Yang Li, Ji-Gang Ren,
  Juan Yin, Qi~Shen, Yuan Cao, Zheng-Ping Li, Feng-Zhi Li, Xia-Wei Chen, Li-Hua
  Sun, Jian-Jun Jia, Jin-Cai Wu, Xiao-Jun Jiang, Jian-Feng Wang, Yong-Mei
  Huang, Qiang Wang, Yi-Lin Zhou, Lei Deng, Tao Xi, Lu~Ma, Tai Hu, Qiang Zhang,
  Yu-Ao Chen, Nai-Le Liu, Xiang-Bin Wang, Zhen-Cai Zhu, Chao-Yang Lu, Rong Shu,
  Cheng-Zhi Peng, Jian-Yu Wang, and Jian-Wei Pan.
\newblock {Satellite-to-ground quantum key distribution}.
\newblock {\em Nature}, 549(7670):43--47, sep 2017.

\bibitem{ride12cqg}
D.~Rideout, T.~Jennewein, G.~Amelino-Camelia, T.~F. Demarie, B.~L. Higgins,
  A.~Kempf, A.~Kent, R.~Laflamme, X.~Ma, R.~B. Mann,
  E.~Mart{\'{i}}n-Mart{\'{i}}nez, N.~C. Menicucci, J.~Moffat, C.~Simon,
  R.~Sorkin, L.~Smolin, and D.~R. Terno.
\newblock {Fundamental quantum optics experiments conceivable with satellites
  reaching relativistic distances and velocities}.
\newblock {\em Classical Quant. Grav.}, 29:224011, 2012.

\bibitem{Safronova2019}
M.~Safronova.
\newblock The search for variation of fundamental constants with clocks.
\newblock {\em Annalen der Physik}, 531:1800364, 2019.

\bibitem{hogan_atom_2015}
Jason~M. Hogan and Mark~A. Kasevich.
\newblock Atom-interferometric gravitational-wave detection using heterodyne
  laser links.
\newblock {\em Phys. Rev. A}, 94:033632, Sep 2016.

\bibitem{Chu1986pro}
S~Chu, A~Ashkin Bjorkholm, P~Gordon, and LW~Hollberg.
\newblock Proposal for optically cooling atoms to temperatures of the order of
  10$^{-6}$\,{K}.
\newblock {\em Opt. Lett.}, 11:73, 1986.

\bibitem{AmmanPRL1997}
Hubert Ammann and Nelson Christensen.
\newblock Delta kick cooling: A new method for cooling atoms.
\newblock {\em Phys. Rev. Lett.}, 78:2088--2091, Mar 1997.

\bibitem{Muentinga13PRL}
H.~M\"untinga, H.~Ahlers, M.~Krutzik, A.~Wenzlawski, S.~Arnold, D.~Becker,
  K.~Bongs, H.~Dittus, H.~Duncker, N.~Gaaloul, C.~Gherasim, E.~Giese,
  C.~Grzeschik, T.~W. H\"ansch, O.~Hellmig, W.~Herr, S.~Herrmann, E.~Kajari,
  S.~Kleinert, C.~L\"ammerzahl, W.~Lewoczko-Adamczyk, J.~Malcolm, N.~Meyer,
  R.~Nolte, A.~Peters, M.~Popp, J.~Reichel, A.~Roura, J.~Rudolph,
  M.~Schiemangk, M.~Schneider, S.~T. Seidel, K.~Sengstock, V.~Tamma,
  T.~Valenzuela, A.~Vogel, R.~Walser, T.~Wendrich, P.~Windpassinger, W.~Zeller,
  T.~van Zoest, W.~Ertmer, W.~P. Schleich, and E.~M. Rasel.
\newblock Interferometry with {Bose}-{Einstein} condensates in microgravity.
\newblock {\em Phys. Rev. Lett.}, 110:093602, Feb 2013.

\bibitem{KovachyPRL2015}
Tim Kovachy, Jason~M. Hogan, Alex Sugarbaker, Susannah~M. Dickerson,
  Christine~A. Donnelly, Chris Overstreet, and Mark~A. Kasevich.
\newblock Matter wave lensing to picokelvin temperatures.
\newblock {\em Phys. Rev. Lett.}, 114:143004, Apr 2015.

\bibitem{KusSte01}
Alexander Kusenko and Paul~J. Steinhardt.
\newblock {Q-ball candidates for self-interacting dark matter}.
\newblock {\em Phys. Rev. Lett.}, 87(14):141301, sep 2001.

\bibitem{Grabowska2018}
Dorota~M Grabowska, Tom Melia, and Surjeet Rajendran.
\newblock {Detecting dark blobs}.
\newblock {\em Physical Review D}, 98(11):115020, dec 2018.

\bibitem{Vilenkin1985}
A.~Vilenkin.
\newblock {Cosmic strings and domain walls}.
\newblock {\em Phys. Rep.}, 121(5):263--315, may 1985.

\bibitem{Vilenkin1994}
A.~Vilenkin and E.~P.~S. Shellard.
\newblock {\em {Cosmic Strings and Other Topological Defects}}.
\newblock Cambridge University Press, Cambridge, England, 1994.

\bibitem{Schneider1999}
P.~Schneider, J.~Ehlers, and E.~Falco.
\newblock {\em {Gravitational Lenses}}.
\newblock Springer, Berlin, Heidelberg, 1999.

\bibitem{Cline2001}
David~B. Cline, editor.
\newblock {\em {Sources and Detection of Dark Matter and Dark Energy in the
  Universe: Fourth International Symposium Held at Marina del Rey, CA, USA
  February 23-25, 2000}}.
\newblock Springer Berlin Heidelberg, Berlin, Heidelberg, 2001.

\bibitem{Ade2014a}
P.~A.~R. Ade, M.~Arnaud, M.~Ashdown, J.~Aumont, C.~Baccigalupi, A.~J. Banday,
  R.~B. Barreiro, E.~Battaner, K.~Benabed, A.~Benoit-L{\'{e}}vy, J.-P. Bernard,
  M.~Bersanelli, P.~Bielewicz, J.~R. Bond, J.~Borrill, F.~R. Bouchet,
  C.~Burigana, J.-F. Cardoso, A.~Catalano, A.~Challinor, A.~Chamballu, H.~C.
  Chiang, P.~R. Christensen, D.~L. Clements, S.~Colombi, L.~P.~L. Colombo,
  F.~Couchot, A.~Coulais, B.~P. Crill, A.~Curto, F.~Cuttaia, L.~Danese, R.~D.
  Davies, R.~J. Davis, P.~de~Bernardis, A.~de~Rosa, G.~de~Zotti,
  J.~Delabrouille, F.-X. D{\'{e}}sert, C.~Dickinson, J.~M. Diego, H.~Dole,
  S.~Donzelli, O.~Dor{\'{e}}, M.~Douspis, X.~Dupac, T.~A. En{\ss}lin, H.~K.
  Eriksen, F.~Finelli, O.~Forni, M.~Frailis, A.~A. Fraisse, E.~Franceschi,
  S.~Galeotta, K.~Ganga, M.~Giard, J.~Gonz{\'{a}}lez-Nuevo, K.~M. G{\'{o}}rski,
  S.~Gratton, A.~Gregorio, A.~Gruppuso, J.~E. Gudmundsson, F.~K. Hansen,
  D.~Hanson, D.~L. Harrison, S.~Henrot-Versill{\'{e}}, D.~Herranz, S.~R.
  Hildebrandt, E.~Hivon, M.~Hobson, W.~A. Holmes, A.~Hornstrup, W.~Hovest,
  K.~M. Huffenberger, A.~H. Jaffe, T.~R. Jaffe, W.~C. Jones, E.~Keih{\"{a}}nen,
  R.~Keskitalo, J.~Knoche, M.~Kunz, H.~Kurki-Suonio, G.~Lagache,
  A.~L{\"{a}}hteenm{\"{a}}ki, J.-M. Lamarre, A.~Lasenby, C.~R. Lawrence,
  R.~Leonardi, J.~Le{\'{o}}n-Tavares, J.~Lesgourgues, M.~Liguori, P.~B. Lilje,
  M.~Linden-V{\o}rnle, M.~L{\'{o}}pez-Caniego, P.~M. Lubin, J.~F.
  Mac{\'{i}}as-P{\'{e}}rez, D.~Maino, N.~Mandolesi, M.~Maris, P.~G. Martin,
  E.~Mart{\'{i}}nez-Gonz{\'{a}}lez, S.~Masi, S.~Matarrese, P.~Mazzotta, P.~R.
  Meinhold, A.~Melchiorri, L.~Mendes, A.~Mennella, M.~Migliaccio, S.~Mitra,
  M.-A. Miville-Desch{\^{e}}nes, A.~Moneti, L.~Montier, G.~Morgante,
  D.~Mortlock, A.~Moss, D.~Munshi, J.~A. Murphy, P.~Naselsky, F.~Nati,
  P.~Natoli, H.~U. N{\o}rgaard-Nielsen, F.~Noviello, D.~Novikov, I.~Novikov,
  C.~A. Oxborrow, L.~Pagano, F.~Pajot, D.~Paoletti, B.~Partridge, F.~Pasian,
  G.~Patanchon, D.~Pearson, T.~J. Pearson, O.~Perdereau, F.~Perrotta,
  F.~Piacentini, M.~Piat, E.~Pierpaoli, D.~Pietrobon, S.~Plaszczynski,
  E.~Pointecouteau, G.~Polenta, N.~Ponthieu, L.~Popa, G.~W. Pratt, S.~Prunet,
  J.-L. Puget, J.~P. Rachen, M.~Reinecke, M.~Remazeilles, C.~Renault,
  S.~Ricciardi, I.~Ristorcelli, G.~Rocha, G.~Roudier, J.~A.
  Rubi{\~{n}}o-Mart{\'{i}}n, B.~Rusholme, M.~Sandri, D.~Scott, V.~Stolyarov,
  R.~Sudiwala, D.~Sutton, A.-S. Suur-Uski, J.-F. Sygnet, J.~A. Tauber,
  L.~Terenzi, L.~Toffolatti, M.~Tomasi, M.~Tristram, M.~Tucci, L.~Valenziano,
  J.~Valiviita, B.~{Van Tent}, P.~Vielva, F.~Villa, L.~A. Wade, B.~D. Wandelt,
  I.~K. Wehus, S.~D.~M. White, D.~Yvon, A.~Zacchei, and A.~Zonca.
\newblock {Planck 2013 results. XXXI. Consistency of the Planck data}.
\newblock {\em Astronomy {\&} Astrophysics}, 571:A31, nov 2014.

\bibitem{Ade2014}
P.~A.~R. Ade, R.~W. Aikin, D.~Barkats, S.~J. Benton, C.~A. Bischoff, J.~J.
  Bock, J.~A. Brevik, I.~Buder, E.~Bullock, C.~D. Dowell, L.~Duband, J.~P.
  Filippini, S.~Fliescher, S.~R. Golwala, M.~Halpern, M.~Hasselfield, S.~R.
  Hildebrandt, G.~C. Hilton, V.~V. Hristov, K.~D. Irwin, K.~S. Karkare, J.~P.
  Kaufman, B.~G. Keating, S.~A. Kernasovskiy, J.~M. Kovac, C.~L. Kuo, E.~M.
  Leitch, M.~Lueker, P.~Mason, C.~B. Netterfield, H.~T. Nguyen, R.~O'Brient,
  R.~W. Ogburn, A.~Orlando, C.~Pryke, C.~D. Reintsema, S.~Richter, R.~Schwarz,
  C.~D. Sheehy, Z.~K. Staniszewski, R.~V. Sudiwala, G.~P. Teply, J.~E. Tolan,
  A.~D. Turner, A.~G. Vieregg, C.~L. Wong, and K.~W. Yoon.
\newblock {Detection of B- Mode Polarization at Degree Angular Scales by
  BICEP2}.
\newblock {\em Phys.\ Rev.\ Lett.}, 112(24):241101, jun 2014.

\bibitem{Moss2014}
Adam Moss and Levon Pogosian.
\newblock {Did BICEP2 See Vector Modes? First $\mathbf{B}$-Mode Constraints on
  Cosmic Defects}.
\newblock {\em Phys. Rev. Lett.}, 112(17):171302, apr 2014.

\bibitem{Freese2013}
K.~Freese, M.~Lisanti, and C.~Savage.
\newblock {Colloquium: Annual modulation of dark matter}.
\newblock {\em Rev. Mod. Phys.}, 85(4):1561, 2013.

\bibitem{Arvanitaki2016}
Asimina Arvanitaki, Peter~W. Graham, Jason~M. Hogan, Surjeet Rajendran, and Ken
  Van~Tilburg.
\newblock Search for light scalar dark matter with atomic gravitational wave
  detectors.
\newblock {\em Phys. Rev. D}, 97:075020, 2018.

\bibitem{Vallone2015}
G.~Vallone, D.~Bacco, D.~Dequal, S.~Gaiarin, V.~Luceri, G.~Bianco, and
  P.~Villoresi.
\newblock {Experimental Satellite Quantum Communications}.
\newblock {\em Phys. Rev. Lett.}, 115:040502, 2015.

\bibitem{pet05}
G.~Petit and P.~Wolf.
\newblock Relativistic theory for time comparisons: a review.
\newblock {\em Metrologia}, 42:S138--S144, 2005.

\bibitem{Graham2013a}
P.~W. Graham, J.~M. Hogan, M.~A. Kasevich, and S.~Rajendran.
\newblock {New Method for Gravitational Wave Detection with Atomic Sensors}.
\newblock {\em Phys. Rev. Lett.}, 110(17):171102, 2013.

\bibitem{SnaddenGG}
M.~J. Snadden, J.~M. McGuirk, P.~Bouyer, K.~G. Haritos, and M.~A. Kasevich.
\newblock {Measurement of the Earth's Gravity Gradient with an Atom
  Interferometer-Based Gravity Gradiometer}.
\newblock {\em Phys. Rev. Lett.}, 81:971--974, 1998.

\bibitem{Sorrentino2014}
F.~Sorrentino, Q.~Bodart, L.~Cacciapuoti, Y.-H. Lien, M.~Prevedelli, G.~Rosi,
  L.~Salvi, and G.~M. Tino.
\newblock {Sensitivity limits of a Raman atom interferometer as a gravity
  gradiometer}.
\newblock {\em Phys. Rev. A}, 89:023607, Feb 2014.

\bibitem{Chiow2016}
S.-w. Chiow, J.~Williams, and N.~Yu.
\newblock {Noise reduction in differential phase extraction of dual atom
  interferometers using an active servo loop}.
\newblock {\em Phys. Rev. A}, 93:013602, Jan 2016.

\bibitem{Yu2011}
N.~Yu and M.~Tinto.
\newblock {Gravitational wave detection with single-laser atom
  interferometers}.
\newblock {\em General Relativity and Gravitation}, 43(7):1943--1952, 2011.

\bibitem{Dimopoulos2008a}
S.~Dimopoulos, P.~Graham, J.~Hogan, and M.~Kasevich.
\newblock {General relativistic effects in atom interferometry}.
\newblock {\em Phys. Rev. D}, 78(4):042003, 2008.

\bibitem{Yu2011conf}
N.~Yu, M.~Tinto, J.~M. Kohel, and R.~J. Thompson.
\newblock {Drag-free atomic acceleration reference for LISA disturbance
  reduction system}.
\newblock In {\em Submission to the NASA RFI: Concepts for the NASA
  Gravitational Wave Mission
  (pcos.gsfc.nasa.gov/studies/gravwave/gravitational-wave-mission-rfis.php)},
  2011.

\bibitem{CHIOW2015}
S.-w. Chiow, J.~Williams, and N.~Yu.
\newblock {Laser-ranging long-baseline differential atom interferometers for
  space}.
\newblock {\em Phys. Rev. A}, 92:063613, 2015.

\bibitem{Geraci2016}
Andrew~A. Geraci and Andrei Derevianko.
\newblock Sensitivity of atom interferometry to ultralight scalar field dark
  matter.
\newblock {\em Phys. Rev. Lett.}, 117:261301, Dec 2016.

\bibitem{Roberts2018a}
B.~M. Roberts, G.~Blewitt, C.~Dailey, and A.~Derevianko.
\newblock {Search for transient ultralight dark matter signatures with networks
  of precision measurement devices using a Bayesian statistics method}.
\newblock {\em Phys. Rev. D}, 97(8):083009, apr 2018.

\bibitem{Olive:2007aj}
Keith~A Olive and Maxim Pospelov.
\newblock {Environmental dependence of masses and coupling constants}.
\newblock {\em Phys. Rev. D}, 77(4):043524, feb 2008.

\bibitem{vall15prl}
G.~Vallone, D.~Bacco, D.~Dequal, S.~Gaiarin, V.~Luceri, G.~Bianco, and
  P.~Villoresi.
\newblock {Experimental Satellite Quantum Communications}.
\newblock {\em Physical Review Letters}, 115:040502, 2015.

\bibitem{vall16prl}
G.~Vallone, D.~Dequal, M.~Tomasin, F.~Vedovato, M.~Schiavon, V.~Luceri,
  G.~Bianco, and P.~Villoresi.
\newblock {Interference at the Single Photon Level Along Satellite-Ground
  Channels}.
\newblock {\em Physical Review Letters}, 116:253601, 2016.

\bibitem{zych12cqg}
M.~Zych, F.~Costa, I.~Pikovski, T.~C. Ralph, and {\v C}.~Brukner.
\newblock {General relativistic effects in quantum interference of photons}.
\newblock {\em Class. Quant. Grav}, 29:224010, 2012.

\bibitem{cole75prl}
R.~Colella, A.~W. Overhauser, and S.~Werner.
\newblock {Observation of gravitationally induced quantum interference}.
\newblock {\em Phys. Rev}, 34, 1975.

\bibitem{marcikic2002pra}
I.~Marcikic, H.~de~Riedmatten, W.~Tittel, V.~Scarani, H.~Zbinden, and N.~Gisin.
\newblock Time-bin entangled qubits for quantum communication created by
  femtosecond pulses.
\newblock {\em Phys. Rev. A}, 66:062308, Dec 2002.

\bibitem{vedovato2018prl}
Francesco Vedovato, Costantino Agnesi, Marco Tomasin, Marco Avesani,
  Jan-\AA{}ke Larsson, Giuseppe Vallone, and Paolo Villoresi.
\newblock Postselection-loophole-free bell violation with genuine time-bin
  entanglement.
\newblock {\em Phys. Rev. Lett.}, 121:190401, Nov 2018.

\bibitem{Poli2014}
N.~Poli, M.~Schioppo, S.~Vogt, St. Falke, U.~Sterr, Ch. Lisdat, and G.~M. Tino.
\newblock {A transportable strontium optical lattice clock}.
\newblock {\em Appl. Phys. B}, 117(4):1107--1116, 2014.

\bibitem{GOCE}
{\em https://www.esa.int/Our\_Activities/Observing\_the\_Earth/GOCE}.

\bibitem{Voyage2050}
{\em https://www.cosmos.esa.int/web/voyage-2050}.

\end{thebibliography}

%
%

\end{document}